\documentclass[prd,preprint,showpacs]{revtex4}
\usepackage{amssymb}
\usepackage{amsmath}
\usepackage{graphicx}
\usepackage{bm}
\usepackage{dcolumn}

\begin{document}

\title{A predictive formulation of the Nambu--Jona-Lasinio
model}
\author{O.A. Battistel$^1$, G. Dallabona$^2$, and G. Krein$^3$ }
\affiliation{$^1$Departamento de F\'{\i}sica, Universidade Federal de Santa
Maria, 97119-900 Santa Maria, RS, Brazil}
\affiliation{$^2$Departamento de Ci\^{e}ncias Exatas, Universidade Federal de Lavras,\\
   Cx. Postal 37, 37200-000, Lavras, MG, Brazil} 
%
\affiliation{$^3$Instituto de F\'{\i}sica Te\'orica, Universidade Estadual
Paulista, Rua Pamplona 145, 01405-900 S\~ao Paulo, SP, Brazil}

\begin{abstract}
A novel strategy to handle divergences typical of perturbative
calculations is implemented for the
Nambu--Jona-Lasinio model and its phenomenological consequences
investigated. The central idea of the method is to avoid the
critical step involved in the regularization process, namely the
explicit evaluation of divergent integrals. This goal is achieved
by assuming a regularization distribution in an implicit way and
making use, in intermediary steps, only of very general properties
of such regularization. The finite parts are separated of the
divergent ones and integrated free from effects of the
regularization. The divergent parts are organized in terms of
standard objects which are independent of the (arbitrary) momenta
running in internal lines of loop graphs. Through the analysis of
symmetry relations, a set of properties for the divergent objects
are identified, which we denominate consistency relations,
reducing the number of divergent objects to only a few ones. 
The calculational strategy
eliminates unphysical dependencies of the arbitrary choices for 
the routing of internal momenta, leading to ambiguity-free, and
symmetry-preserving physical amplitudes. 
We show that the imposition of scale properties for the basic
divergent objects leads to a critical condition for the constituent
quark mass such that the remaining arbitrariness is removed.
The model become predictive in the sense that its phenomenological
consequences do not depend on possible choices made in intermediary
steps. Numerical results are
obtained for physical quantities at the one-loop level for the pion and
sigma masses and pion-quark and sigma-quark coupling constants.
\end{abstract}

\maketitle

\section{Introduction}

It is largely accepted that many of the essential features of
chiral symmetry in quantum chromodynamics (QCD) are captured by
the Nambu--Jona-Lasinio (NJL) model~\cite{Nambu}, a simple
relativistic quantum field theory with (nonrenormalizable)
four-fermion interactions. In~the limit of exact chiral symmetry
the fermions are massless and the interaction Lagrangian density
of the model, in its simplest version, contains the chirally symmetric 
sum of scalar and pseudo-scalar four-fermion interactions. Since the
first works, using the model with quark degrees of freedom, in the
earlier days of QCD~\cite{Eguchi, Kikkawa}, the model has been
extensively used to describe low energy hadronic observables, like
hadronic masses, correlation and structure functions in vacuum
and at finite densities and temperatures -- for a complete list of
references see the reviews in Refs.~\cite{VW, Klevansky, HK-rev,
Bochum, Bijnens, Bernard, Ripka, Bub}.

One of the reasons for the widespread use of the model is that it
realizes the dynamical breaking of chiral symmetry already at the
one-loop (mean field) approximation. 
The predictions of the model, however, are intimately
compromising with the specific strategy adopted to handle the
ultraviolet divergences given the nonrenormalizable nature of the model. 
As a consequence, the
specification of a procedure for handling divergent amplitudes is
a necessary and essential first step to be performed before extracting physical
predictions as must be made in any relativistic quantum field theory. In a
renormalizable theory this is done by specifying a regularization
procedure by which the divergences are isolated and eliminated
through a convenient reparametrization of the theory, removing, in
this way, any parameters introduced in the regularization process.
Therefore, although essential and necessary, the regularization
process plays a secondary role and seems to be a disposable
intermediate step, in the sense that it is not meant to modify the
physical content of the theory. However, there is a distinctive and
nontrivial aspect with the regularization of the ultraviolet
divergences in the NJL model in view of its nonrenormalizability.
Unlike with renormalizable models, as an increasing number of
loops is considered the reparametrization of the model can be made
only at the cost of adding an increasing number of terms with extra 
coupling constants to the original Lagrangian in order to render physical 
amplitudes independent of the regularization procedure. In principle, 
there is no problem with a theory having an infinite number of coupling 
constants when using it as an effective field theory, as explained by 
Weinberg~\cite{Weinberg}. However, practitioners of the NJL model have 
followed the attitude of using it as a regularization-dependent model, 
considering the regularization procedure part of the definition of the 
model. Within such an attitude a large body of interesting and valuable
work has been, and continues to be done using the model.

The regularization of divergent amplitudes is a delicate process
due to the arbitrariness in the manipulation of improper integrals
which can be converted into ambiguities when results become
dependent on the choices involved. There are ambiguities
associated with the arbitrary routing of the momenta in internal
lines of divergent loop amplitudes, which invariably lead to the
violation of space-time homogeneity. There are also ambiguities
associated with the choice of the common scale for the divergent
and finite parts of amplitudes that may lead to the breaking of
scale invariance. In general, different sorts of ambiguities have
the potential of leading to violations of symmetry relations of
global and local gauge symmetries. The most commonly used
regularization procedures for the NJL model such as three- and
four-momentum cutoff, Pauli-Villars and proper-time lead to one or
more of such symmetry violations. Dimensional regularization (DR), although 
not much used within the NJL model, in general, leads to amplitudes free 
from ambiguities and symmetry preserving. However, it has problems at 
high densities and temperatures, when chiral symmetry is restored. 
This is due to the fact that the quadratic divergence which appears 
in almost all one-loop amplitudes must be assumed as zero in the zero-mass limit. 
Practice with the NJL model has shown that depending on the problem 
studied, one regularization scheme seems to be more appropriate than 
another because of the problems just mentioned. For example, when 
working with correlation functions, in general dispersion relations 
are not automatically fulfilled in cutoff and proper-time
regularizations, in contrast to Pauli-Villars regularization. On the
other hand, while causality is preserved with Pauli-Villars 
regularization, unitarity is violated at high enough energies, 
although it is preserved with proper-time regularization. These sorts 
of problems are well known and arguments have been put forward and tricks
invented to deal with such problems -- for a discussion on
these issues, see for example Refs.~\cite{Bron,Doring,DavERA}.
Obviously, this situation is unsatisfactory since one would like
that the regularization scheme play a secondary role in the process
of making predictions with the model.

The difficulties pointed out above lead Willey~\cite{Willey} and
Gherghetta~\cite{Gherghetta} to conclude that there is no way to
make consistent physical predictions with the NJL model using
traditional regularization techniques. However, this question in
the context of the gauged NJL model was considered in a later work
by Battistel and Nemes~\cite{Orimar-Nemes} using a novel strategy
to handle divergent amplitudes~\cite{Orimar-tese}. The referred
investigation revealed that NJL amplitudes ambiguity-free and
symmetry-preserving can be obtained, and as such making the NJL
model predictive. The central idea of the method is to avoid the
critical step involved in the regularization process, namely the
explicit evaluation of divergent integrals. This goal is achieved
by assuming a regularization distribution in an implicit way and
making use, in intermediary steps, only of very general properties
of such regularization. The finite parts are separated of the
divergent ones and integrated free from effects of the
regularization in a completely similar way as made in the treatment
of renormalizable theories. The divergent parts are organized in terms of
standard objects which are independent of the (arbitrary) momenta
running in internal lines of loop graphs. Through the analysis of
symmetry relations, a set of properties for the divergent objects
are identified, which we denominate consistency relations (CR's),
reducing the number of divergent objects to only a few ones. The
remaining objects never really need to be evaluated. In
renormalizable theories they are eliminated by the counter-terms.
In a non-renormalizable model, such as in the NJL, the basic
divergences are fixed by fitting observables, as we will see along
this contribution.

Having in mind this perspective, in the present work we extend the
original discussion of Ref.~\cite{Orimar-Nemes} by presenting a
complete and unified discussion on the symmetry relations involving
Green's functions, including tensor operators. Some of the
relations can be derived using the methods of current algebra, 
in particular when using the conservation of the fermionic vector 
current and the proportionality of the divergence of the axial 
vector current to the pseudoscalar current. Here we also discuss 
relations of Green's functions of tensor operators. Such relations
cannot be obtained through current algebra methods, because 
the divergence of the fermionic tensor current cannot be written in 
terms of other fermionic currents. This is the case, for example, of 
the relations involving tensor-tensor two-point amplitudes. In a 
first step of our investigation we will show that it is possible to
obtain physical amplitudes preserving the symmetries and
automatically free from ambiguities associated with the
arbitrariness in the routing of momenta in the internal lines of
loops. In a second step we will show that the undefined quantities
associated with the divergent objects can be fixed
phenomenologically, without to the recourse of calculating any
divergent integral, leading in this way to a regularization
independent parametrization of the model. The model, within this
formulation, becomes predictive in the sense that all
arbitrariness are removed from the physical amplitudes. The model
works like a ``renormalizable'' theory at the one-loop level.

The results obtained in the present paper are new and extend the
applicability of the NJL model in way that it becomes independent
of a particular regularization scheme, since no explicit
regularization is actually used. It is new because all symmetry
constraints on general Green's functions, including tensor ones,
are preserved and as such no problems with causality and unitarity
can arise. The present method has also been applied in the context
of CPT breaking in models with Chern-Simons interactions
\cite{Orimar-cpt}, to the neutral electromagnetic pion-decay where
the $AVV$ triangle anomaly phenomenon is discussed, triangle
anomalies~\cite{Orimar-Orildo} and tensor densities~\cite{Gilson}.
In particular, it was shown that the adopted strategy furnish the expected
anomalous term and the ambiguities again play no relevant role
\cite{Orimar-pion}. One important aspect of the method presented
here is that, since no explicit regulator is used, a result
obtained within a given traditional regularization method can be
immediately reobtained by explicit evaluation of the implicitly 
regulated expressions. With this, the use of the consistency relations 
allow to identify the reasons why commonly used regularization schemes 
lead to symmetry violations. Invariably, the reasons are that not all 
the consistency relations are satisfied simultaneously within the 
traditional regularization methods.

The plan of this paper is the following. In
Section~\ref{sec:consistency}, we present the Lagrangian density
of the NJL model used in this paper, and discuss general
``consistency constraints" that the one-loop Green's functions
must satisfy in order not to violate symmetry relations. Next, in
Section~\ref{sec:isolating} we discuss a calculational scheme that
preserves the general relationships among the Greens functions
obtained in Section~\ref{sec:consistency}. The calculational
scheme isolates the purely divergent terms, which will disappear
because of symmetry consideration or will be fitted to
observables, while the finite parts are integrated without any
regularization. In section~\ref{sec:ambiguities}, we discuss the
ambiguities associated with the choices for the momentum routing
in the internal lines of loops and show that the methods used to
isolate the divergent parts respects all the general relations
among Green's functions. This aspect of the regularization is
central to the paper and is highly nontrivial since,  because of
the ultraviolet divergences, these relations can very easily be
violated when not being careful with the explicit evaluation of
the integrals within a particular regularization scheme. In
Section~\ref{sec:symmetries} we identify the general properties
that the divergent ambiguous quantities must satisfy in order to
guarantee the preservation of fundamental symmetries. For that,
Ward identities and other general constraints imposed by Furry's
theorem will be invoked. The phenomenology and numerical results
are presented in Section~\ref{sec:numerical}. Here only the
traditional observables, like pion and sigma masses and coupling
constants are calculated to show that the traditional
phenomenology is obtained in a straightforward way. Our Conclusions
and Perspectives for future work are presented in
Section~\ref{sec:conclusions}.

\section{Model Lagrangian and consistency constraints}
\label{sec:consistency}

In this paper we restrict the discussion to the simplest SU(2)
version of the NJL model that incorporates the light-quark $u$ and
$d$ flavors only. The SU(3) case will be considered elsewhere \cite{OB_su3}. 
The Lagrangian density is given by
\begin{eqnarray}
\mathcal{L} &=&\bar{\psi}\left( i\not\!{\partial}-m_{0}\right) \psi + G_{S}
\left[ \left( \bar{\psi}\psi \right) ^{2} + \left(
\bar{\psi}\vec{\tau}i\gamma _{5}\psi \right) ^{2}\right]  ,
\label{lagrangian}
\end{eqnarray}
where $\psi $ is the two-flavor, three-color quark field operator
and $m_{0}$ is the diagonal current quark mass matrix. To simplify
the discussion, we take equal $u$ and $d$ quark masses. The
nonperturbative quark propagator $S(p)$ is given in terms of the
self-energy $\Sigma(p)$ as
\begin{equation}
S^{-1}(p) = \not\!p - \Sigma(p). \label{nonpS}
\end{equation}
In the mean field approximation, the self-energy is momentum
independent $\Sigma(p) \equiv M$, with $M$ satisfying a gap
equation~\cite{Nambu}
\begin{equation}
M = m_{0} - 2 \, G_{S}N_f \left\langle \overline{\psi }\psi
\right\rangle , \label{gap1}
\end{equation}
where $N_f = 2$ is the number of flavors, and $\left\langle
\overline{\psi }\psi \right\rangle $ is the one-flavor, Lorentz
scalar one-point function (the quark condensate) given by
\begin{equation}
\left\langle \overline{\psi }\psi \right\rangle =  - i \int
\frac{d^4k}{(2\pi)^4} {\rm Tr} [S(k)] = - 4 N_{c} \, i \int
\frac{d^4k}{(2\pi)^4} \frac{M}{k^2 - M^2} , \label{cond}
\end{equation}
where $N_c = 3$ is the number of colors.

In general, phenomenological predictions for meson masses and
correlation functions require the evaluation of purely fermionic
$n$-point Green's functions. In spite the fact the model is
nonrenormalizable, the Green's functions obey well-defined
relations among them. Such relations are the manifestation of the
symmetries of the underlying Lagrangian defining the model.
Therefore, in any attempt of describing a specific phenomenology
it is crucial that the evaluation of physical amplitudes preserve
such symmetry relations. If it turns that such symmetry relations
are not preserved by the calculation, the predictions cannot be
characterized as consequences of the underlying symmetries
supposed relevant for the specific phenomenology and which were
the main motivation for using a schematic model like the NJL
model. 

A generic one-loop $n$-point Green's function can be defined as
\begin{equation}
T^{\Gamma _{1}\Gamma _{2}\cdots \Gamma _{n}}\left(
k_{1},k_{2},\cdots ,k_{n}\right) = \int \frac{d^{4}k}{(2\pi
)^{4}}\mathrm{Tr}\left[ \Gamma _{1}S\left( k+k_{1}\right) \Gamma_2
S\left( k+k_{2}\right) \cdots \Gamma _{n}S\left( k+k_{n}\right)
\right] , \label{defT}
\end{equation}
where the trace is over Dirac indices only, and the $\Gamma_i$
represent one or more of the matrices $\left(1, \gamma_{\mu},
\gamma _{5}, \gamma_{\mu} \gamma_{5}, \sigma_{\mu \nu} \right)$ to
which we attribute the labels $\left( S,V,P,A,T\right) $,
respectively. The fermion propagators are given by
Eq.~(\ref{nonpS}) and the $k_{n}$ are arbitrary routing momenta in
the internal lines, and are related to the external momenta. A
physical amplitude can depend only on differences of $k_n$, any
dependence on sums of $k_n$ is unphysical.  We note that the
highest superficial degree of divergence is cubic, and occurs for
the amplitude with $n=1$. For $n > 4$, the amplitudes are finite.
In particular, in the NJL model the scalar one-point function is
relevant for the gap equation, the two-point functions appear in
the bound-state equations for mesons, the amplitudes with $n=3$
describe meson-decays and the four-point functions are relevant
for meson-meson scattering.

At the one loop level, there appear only two divergences, a
quadratic and the logarithmic. 
The cubic divergence is absent in the one-point functions either
because of the trace or because the integral is identically zero -
for the same reason there is no linear divergence in two- and
three-point functions. The standard procedure to deal with the
divergences is to cutoff the momentum integrals at some momentum
$\Lambda$. With this, the model has two unknowns, the coupling
$G_S$ and $\Lambda$. These can be fitted by using the values of
the quark condensate $\langle\overline{\psi}\psi\rangle$, related
to the quadratically divergent scalar one-point function $T^S$,
and the pion decay constant $f_\pi$, related to the logarithmically
divergent axial-vector pseudo-scalar two-point function $T^{AP}$.
Since these two types of divergences are the only ones that appear
in all other Green's functions of the model, all divergences can
be absorbed by the physical quantities
$\langle\overline{\psi}\psi\rangle$ and  $f_\pi$. In a certain
sense, this is a type of renormalization.

There are two aspects we would like to note with respect to this
``renormalization''. First, one is explicitly using relations among
divergent Green's functions having different numbers of points.
Second, given the strict nonrenormalizability of the model,
divergent amplitudes are related to physical quantities through a
regularization function or, in last instance, through the
adjustment of regularization parameters which are interpreted as
cutoffs in the momentum integration. It is well known that these
two facts lead in general to symmetry violations. Moreover, the
effects of the modification introduced in the regularization
process remain present even in the finite parts of the amplitudes.
In this respect, the regularization and renormalization procedure
for nonrenormalizable and renormalizable models are treated in
completely different manner. This discussion emphasizes the
difficulties in satisfying kinematical constraints and symmetry
relations involving one or more divergent amplitudes. The method
we present here is based on the following strategy: (1) first, all
constraints and symmetry relations are imposed without evaluating
any divergent integral, (2) divergences are isolated into
quantities that are independent of arbitrary routing momenta, and
(3) amplitudes respecting all symmetry relations and free
from ambiguities are obtained. All this is achieved without
compromising with a particular regularization scheme.

Let us start considering  the general \textquotedblleft
consistency constraints\textquotedblright that Green's functions
must satisfy in order not to violate symmetry relations. One very
general and powerful way to generate relations among Green's
functions is to identify identities, at the integrand level,
resulting from the contraction of the Lorentz vector indices of a
vertex operator with an external momentum. Although this method is
entirely equivalent to the current algebra technique for some
types of amplitudes, it can be applied also to tensor currents --
for which the methods of current algebra are not applicable. As an
example consider the identity
\begin{equation}
\left( k_{1}-k_{2}\right) ^{\nu }\left[ \gamma _{\mu }S\left( k+k_{1}\right)
\gamma _{\nu }S\left( k+k_{2}\right) \right] =\gamma _{\mu }S\left(
k+k_{2}\right) -\gamma _{\mu }S\left( k+k_{1}\right) ,
\end{equation}
where $S(k)$ is the mean field quark operator given by
Eq.~(\ref{nonpS}). This identity follows trivially from the
algebra of the Dirac $\gamma$ matrices. Tracing both sides and
integrating in momentum, a genuine relation among Green's
functions of the model is obtained
\begin{equation}
\left( k_{1}-k_{2}\right) ^{\nu }T_{\mu \nu }^{VV}\left(
k_{1},k_{2}\right) = T_{\mu }^{V}(k_{1})-T_{\mu }^{V}(k_{2}),
\label{rel1-T-VV}
\end{equation}
where $T_{\mu }^{V}$ and $T_{\mu \nu }^{VV}$ are respectively the
vector one-point function ($\Gamma_1 = \gamma_\mu$) and the
two-point vector-vector amplitudes ($\Gamma_1 = \gamma_\mu,
\Gamma_2 = \gamma_\nu)$ - see Eq.~(\ref{defT}). In the same way,
one obtains
\begin{equation}
\left( k_{1}-k_{2}\right) ^{\mu }T_{\mu \nu }^{VV}\left(
k_{1},k_{2}\right) = T_{\nu }^{V}(k_{1})-T_{\nu }^{V}(k_{2}).
\label{rel-T-VV}
\end{equation}
Similarly, one can contract a Lorentz axial-vector two-point
density with an external momentum as
\begin{eqnarray}
(k_{1}-k_{2})^{\mu }\left[ \gamma _{\nu } \gamma _{5} S(k+k_{1})
\gamma _{\mu} \gamma_{5} S(k+k_{2}) \right]  &=& 2 M \left[
\gamma_{\nu } \gamma _{5}S(k+k_{1})\gamma _{5}S(k+k_{2})\right]
\nonumber \\
&+& \gamma_{\nu} S(k+k_{2}) + \gamma_{\nu }\gamma_{5} S(k+k_{1})
\gamma_{5},
\end{eqnarray}
where we have used the anticommutation of $\gamma_{5}$ and the
$\gamma _{\nu }$ matrices. Again, taking the traces and
integrating on both sides, we obtain
\begin{equation}
(k_{1}-k_{2})^{\mu } T_{\nu \mu }^{AA}\left( k_{1},k_{2}\right) =
2M T_{\nu }^{AP}\left( k_{1},k_{2}\right) +T_{\nu
}^{V}(k_{2})-T_{\nu }^{V}(k_{1}), \label{rel-T-AA}
\end{equation}
where the indices $A$ and $P$ stand for axial and pseudoscalar
corresponding respectively to $\Gamma = \gamma_\mu\gamma_5$ and
$\Gamma = \gamma_5$, as defined in the paragraph following
Eq.~(\ref{defT}). Following the procedure described above, we can
get also the relations
\begin{eqnarray}
(k_{1}-k_{2})^{\nu }T_{\nu }^{VS}\left( k_{1},k_{2}\right)
&=&T^{S}(k_{2})-T^{S}(k_{1}),  \label{rel-T-VS} \\
(k_{1}-k_{2})^{\nu }T_{\nu }^{AP}\left( k_{1},k_{2}\right)
&=&-2MT^{PP}\left( k_{1},k_{2}\right) +T^{S}(k_{2})-T^{S}(k_{1}),
\label{rel-T-AP} \\
(k_{1}-k_{2})^{\mu }T_{\mu \nu }^{AV}\left( k_{1},k_{2}\right) &=&
- 2M T_{\nu }^{PV}\left( k_{1},k_{2}\right) + T_{\nu}^{A}(k_{2}) -
T_{\nu }^{A}(k_{1}), \label{rel-T-AV1} \\
(k_{1}-k_{2})^{\nu }T_{\mu \nu }^{AV}\left( k_{1},k_{2}\right) &=&
T_{\nu}^{A}(k_{2}) - T_{\nu }^{A}(k_{1}).  \label{rel-T-AV2}
\end{eqnarray}
The expressions (\ref{rel-T-VS})-(\ref{rel-T-AV2}) are nothing more 
than the relations that follow
from current algebra methods when using the conservation of the
fermionic vector current and the proportionality of the divergence
of the axial vector current with the pseudoscalar current. The
expressions involving the one-point functions correspond to the
current commutator terms.
Additional relations can be identified at the trace level
\begin{eqnarray}
&& T_{\mu \nu }^{AV}\left( k_{1},k_{2}\right) = \frac{i}{2M}
\varepsilon_{\mu \nu \lambda \sigma } \left(
k_{1}-k_{2}\right)^{\lambda} \left( T^{SV}\right)^{\sigma}
\left(k_{1},k_{2}\right) ,  \label{ident1} \\
&& T_{\mu }^{AP}\left( k_{1},k_{2}\right) = -\frac{1}{2M}\left(
k_{1}-k_{2}\right) _{\mu }\left[ T^{SS}\left( k_{1},k_{2}\right)
+ T^{PP}\left( k_{1},k_{2}\right) \right] , \\
&& T_{\mu \nu }^{VV}\left( k_{1},k_{2}\right) - T_{\mu \nu
}^{AA}\left( k_{1},k_{2}\right) = g_{\mu \nu }\left[ T^{SS}\left(
k_{1},k_{2}\right) +T^{PP}\left( k_{1},k_{2}\right) \right] .
\label{ident3}
\end{eqnarray}

One of the advantages of the method used above is that it can be
used to obtain relations for Green's functions involving tensor
operators. Such relations cannot be obtained through current
algebra methods simply because the divergence of the fermionic
tensor current cannot be written in terms of other fermionic
currents. This is the case of relations involving tensor-tensor
two-point amplitudes, like for example the following one
\begin{eqnarray}
\left( k_{1}-k_{2}\right)^{\mu }T_{\mu \nu \alpha \beta }^{TT}
\left(k_{1},k_{2}\right) &=& - g_{\alpha \nu }\left[ T_{\beta
}^{V}\left( k_{2}\right) - T_{\beta }^{V}\left( k_{1}\right)
+\left( k_{1}-k_{2}\right)_{\beta } T^{SS}\left(
k_{1},k_{2}\right) \right]  \nonumber \\
&&+ g_{\beta \nu }\left[ T_{\alpha }^{V}\left( k_{2}\right) -
T_{\alpha }^{V}\left( k_{1}\right) +\left( k_{1}-k_{2}\right)
_{\alpha} T^{SS}\left(k_{1},k_{2}\right) \right]  \nonumber \\
&&+\left( k_{1}-k_{2}\right) _{\alpha }T_{\nu \beta }^{AA}
\left(k_{1},k_{2}\right) - \left( k_{1}-k_{2}\right)_{\beta}
T_{\nu\alpha}^{AA} \left( k_{1},k_{2}\right)  \nonumber \\
&&- 2M T_{\nu \alpha \beta }^{VT}\left( k_{1},k_{2}\right) .
\label{rel-T-TT}
\end{eqnarray}
Given the fact that three other Lorentz indexes are left
uncontracted in the $TT$ two-point function, it is also possible
to establish constraints on successive contractions with the
external momenta. It is immediate to note that when these
contractions involve both Lorentz indexes of a tensor operator,
the result must vanish identically
\begin{equation}
\left(k_{1}-k_{2}\right)^{\alpha } \left(
k_{1}-k_{2}\right)^{\beta} T_{\mu\nu\alpha\beta}^{TT}
\left(k_{1},k_{2}\right) = 0 .
\end{equation}
This is due to the fact that $T_{\mu\nu\alpha\beta}^{TT} \sim
\sigma_{\alpha \beta}$ , where $ \sigma_{\alpha\beta} = - i/2 \,
[\gamma_\alpha, \gamma_\beta]$, and therefore $\left( k_{1}-k_{2}
\right)^{\alpha} \left( k_{1}-k_{2} \right)^{\beta }
\sigma_{\alpha \beta}=0$. This property imposes additional
constraints on the consistent evaluation of $TT$ two-point
function. Even if such requirements seem to be obvious at this
point, the divergent character of the integrals defining the
amplitude makes satisfaction of this property far from being
trivial.

Following strictly the same procedure outlined above, relations
involving the remaining tensor two-point functions can be
established as
\begin{eqnarray}
\left( k_{1}-k_{2}\right) ^{\alpha }T_{\mu \nu \alpha }^{TV}\left(
k_{1},k_{2}\right) &=& 0,  \label{rel-T-VT1} \\
\left( k_{1}-k_{2}\right) ^{\mu }T_{\mu \nu }^{TP}\left(
k_{1},k_{2}\right) &=& -2M T_{\nu }^{VP}\left( k_{1},k_{2}\right)
, \label{rel-T-TP} \end{eqnarray}
\begin{eqnarray}
\left( k_{1}-k_{2}\right)^{\alpha }T_{\mu \nu \alpha }^{TA} \left(
k_{1},k_{2}\right) &=& 2M T_{\mu \nu }^{TP} \left(
k_{1},k_{2}\right) ,
\label{rel-T-TA1}
\end{eqnarray}
\begin{eqnarray}
\left( k_{1}-k_{2}\right) ^{\mu }T_{\mu \nu \alpha }^{TA}\left(
k_{1},k_{2}\right) &=& - 2M T_{\nu \alpha }^{VA} \left(
k_{1},k_{2} \right) = -i\varepsilon _{\nu \alpha \lambda \xi
}\left( k_{1}-k_{2}\right) ^{\lambda }\left( T^{SV}\right) ^{\xi
}\left(k_{1},k_{2}\right) ,
\label{rel-T-TA2}
\end{eqnarray}
\begin{eqnarray}
\left( k_{1}-k_{2}\right) ^{\mu }T_{\mu \nu }^{TS}\left(
k_{1},k_{2}\right) &=&\frac{1}{2m}\left\{ \left(
k_{1}-k_{2}\right) ^{2}T_{\nu }^{VS}\left( k_{1},k_{2}\right)
-\left( k_{1}-k_{2}\right) _{\nu }\left[ T^{S}\left( k_{2}\right)
-T^{S}\left( k_{1}\right) \right] \right\} \nonumber \\
&=& - \frac{i}{2} \varepsilon_{\alpha \nu \lambda \xi }\left(
k_{1}-k_{2}\right) ^{\alpha }\left( T^{AV}\right)^{\lambda\xi}
\left(k_{1},k_{2}\right) ,  \label{rel-T-TS} 
\end{eqnarray}
\begin{eqnarray}
\left( k_{1}-k_{2}\right) ^{\mu }T_{\mu \nu \alpha }^{TV}\left(
k_{1},k_{2}\right) &=&\frac{1}{2m}\left( k_{1}-k_{2}\right)
^{2}\left[ T_{\alpha \nu }^{VV}\left( k_{1},k_{2}\right)
-T_{\alpha \nu }^{AA}\left( k_{1},k_{2}\right) \right] \nonumber \\
&&- \left( k_{1} - k_{2} \right)_{\nu} T_{\alpha }^{PA} \left(
k_{1},k_{2}\right) .  \label{rel-T-VT2}
\end{eqnarray}
In deriving these results, we have used the relations given in
Eqs.~(\ref{ident1})-(\ref{ident3}).

At this point it is very important to note that the results
obtained above for the contraction of amplitudes with the external
momentum are of general validity. Only algebraic manipulations
have been made, no single divergent integral has been evaluated
and no changes of variables under integration have been made.
Thus, the results obtained are not compromised with any type of
regularization. The important question one has to face when
evaluating the integrals defining the different amplitudes, is how
to give a meaning to the divergent integrals without violating
these relations. This will be discussed in the next section.

\section{Calculational Scheme for Handling Divergent Integrals}
\label{sec:isolating}

In the preceding section we have considered the Green's functions
for which one needs to construct a calculational scheme that
preserves the general relationships among them, which we
denominated consistency constraints. As a matter of consistency,
all the constraints must be fulfilled without specifying special
choices of the loop momenta. Divergences will appear in the form
of improper Feynman integrals, but only a small number of them
need to be evaluated since all the amplitudes are combinations of
a few Feynman integrals.

The traditional way to handle a divergent Feynman integral is to
adopt an explicit regularization. Invariably, this amounts in
modifying the original integral in a way the integration becomes
well defined. In the context of NJL models, such modifications are
commonly made by introducing in the integrand a distribution in
the loop momentum to render the integral convergent. In doing so,
the results of the integrals become a function of the parameters
of the regulating distribution. In perturbatively renormalizable
theories, one tries to isolate the purely divergent terms in order
to specify the adequate counterterms for the renormalization. The
parts which are independent of the regulating parameters are
identified as the finite parts and carry the physical content of
the amplitude. This means that the functions of the physical
momenta are not affected by the regularization in the limit the
regularization is removed. The procedure we adopt to use for the
NJL model follows closely this general strategy,  the purely
divergent parts of the amplitudes will be fitted to observables,
while the finite parts are integrated without any regularization.

In the first step one assumes an unspecified regularization
distribution $G_{\Lambda_i} \left( k^{2},\Lambda^2_i \right)$
dependent on one or more ``cutoff"  parameters $\Lambda_{i}$, such
that the original divergent integral is replaced by a finite one
as~\cite{Orimar-tese}
\begin{equation}
\int \frac{d^{4}k}{\left( 2\pi \right)^{4}}f(k) \rightarrow \int
\frac{d^4 k}{(2\pi)^4} f(k) \, G_{\Lambda_i} \left(
k^{2},\Lambda^2_i \right) \equiv \int_{\Lambda }
\frac{d^4k}{(2\pi)^4} f(k)  .
\end{equation}
The generic distribution $G(k^{2},\Lambda _{i}^{2})$, in addition
of having the obvious property of turning the original integral
convergent, must depend only on $k^2$ due to Lorentz invariance
and must have the limit
\begin{equation}
\lim_{\Lambda _{i}^{2}\rightarrow \infty }G_{\Lambda _{i}}\left(
k^{2},\Lambda _{i}^{2}\right) =1. \label{norm}
\end{equation}
which allows us to connect the regularized integral with
the original one.
Having assumed the existence of such a regularization
distribution, one manipulates the integrand of the divergent
integral in a way to isolate all the divergences in
momentum-independent integrals. This goal can be achieved by using
the identity
\begin{equation}
\frac{1}{[(k+k_{i})^{2} - M^{2}]} = \sum_{j=0}^{N}\frac{\left(
-1\right) ^{j}\left( k_{i}^{2}+2k_{i}\cdot k\right) ^{j}}{\left(
k^{2} - M^{2}\right)^{j+1}} + \frac{\left( -1\right)^{N+1}\left(
k_{i}^{2}+2k_{i}\cdot k\right) ^{N+1}}{\left( k^{2}-M^{2}\right)
^{N+1}\left[ \left( k+k_{i}\right) ^{2}-M^{2}\right] },
\end{equation}
where the $k_{i}$ is (in principle) an arbitrary routing momentum
of an internal line in a loop, and $M$ is the fermion mass running
in the loop. The value of the integer $N$ is the smallest integer
that makes integrals involving this last term finite when removing
the regulating function. In view of Eq.~(\ref{norm}), the
corresponding integration can be performed without restrictions
and will be free from the specific effects of an eventual
regularization. No additional assumptions are made with respect to
the remaining divergent terms.

When evaluating the amplitudes considered in the previous section,
after taking the appropriate traces, it is not difficult to
convince ourselves that only five divergent Feynman integrals will
appear, namely
\begin{equation}
\left( I_{1};I_{1}^{\mu }\right) =\int \frac{d^{4}k}{(2\pi )^{4}}\frac{
\left( 1;k^{\mu }\right) }{\left[ \left( k+k_{1}\right)
^{2}-M^{2}\right] }, \label{I1}
\end{equation}
\begin{equation}
\left( I_{2};I_{2}^{\mu };I_{2}^{\mu \nu }\right) =\int
\frac{d^{4}k}{(2\pi )^{4}}\frac{\left( 1;k^{\mu };k^{\mu }k^{\nu
}\right) }{\left[ \left(
k+k_{1}\right) ^{2}-M^{2}\right] \left[ \left( k+k_{2}\right) ^{2}-M^{2}
\right] }.  \label{I2}
\end{equation}
The divergent parts of these integrals can be rewritten in terms
of five divergent quantities that we denote by $\square _{\alpha
\beta \mu \nu }$, $\Delta _{\mu \nu }$, $\nabla_{\mu\nu}$,
$I_{log}$ and $I_{quad}$. Specifically, the integrals $I_{1}$ and
$I_{1\mu}$ that appear in the one-point amplitudes can be written
as
\begin{eqnarray}
I_{1}\left( k_{1}\right) &=& \left[ I_{quad}(M^{2})\right]
+k_{1}^{\alpha }k_{1}^{\beta }\left[ \Delta _{\alpha \beta
}\right] , \\
I_{1\mu }\left( k_{1}\right) &=& -k_{1\mu }\left[
I_{quad}(M^{2})\right] -k_{1}^{\beta }\left[ \nabla _{\beta \mu
}\right] -\frac{1}{3}k_{1}^{\beta }k_{1}^{\alpha }k_{1}^{\nu
}\left[ \square _{\alpha \beta \mu \nu }\right] \nonumber \\
&-& \frac{1}{3}k_{1\mu }k_{1}^{\alpha }k_{1}^{\beta }\left[ \Delta
_{\alpha \beta }\right] +\frac{1}{3}k_{1}^{2}k_{1}^{\alpha }\left[
\Delta _{\alpha \mu }\right] .
\end{eqnarray}
The integrals $I_2$, $I_{2\mu}$ and $I_{2\mu\nu}$ related to
two-point amplitudes can be written as
\begin{eqnarray}
I_{2}\left( k_{1},k_{2}\right) &=& \left[ I_{\log }\left(
M^{2}\right) \right] -i(4\pi )^{-2}\left[ Z_{0}\left( \left(
k_{1}-k_{2}\right) ^{2};M^{2}\right) \right] \\
I_{2\mu }\left( k_{1},k_{2}\right) &=&-\frac{1}{2}(k_{1}+k_{2})^{\alpha }
\left[ \Delta _{\alpha \mu }\right] -\frac{1}{2}(k_{1}+k_{2})_{\mu
}\left( I_{2}\right) , \\
I_{2\mu \nu }\left( k_{1},k_{2}\right) &=&\frac{1}{2}\left[ \nabla
_{\mu \nu
}\right] -\frac{1}{12}\left( k_{1}-k_{2}\right) ^{2}\left[ \Delta _{\mu \nu }
\right]  \notag \\
&&+\frac{1}{6}\left( k_{2}^{\alpha }k_{2}^{\beta }+k_{1}^{\alpha
}k_{2}^{\beta } + k_{1}^{\alpha }k_{1}^{\beta }\right) \left[
\square _{\alpha
\beta \mu \nu }\right]  \notag \\
&& + \frac{1}{6}\left( k_{2\nu }k_{2}^{\beta } + k_{1\nu}
k_{2}^{\beta } + k_{1\nu} k_{1}^{\beta }\right) \left[ \Delta
_{\beta \mu }\right]
\notag \\
&&+\frac{1}{6}\left( k_{2\mu }k_{2}^{\beta }+k_{1\mu }k_{2}^{\beta
}+k_{1\mu
}k_{1}^{\beta }\right) \left[ \Delta _{\beta \nu }\right]  \notag \\
&&+\frac{1}{2}g_{\mu \nu }\left[ I_{quad}\left( M^{2}\right) \right] -\frac{1
}{12}g_{\mu \nu }\left( k_{1}-k_{2}\right) ^{2}\left[ I_{\log
}\left(
M^{2}\right) \right]  \notag \\
&&+\frac{1}{6}\left( 2k_{2\nu }k_{2\mu }+k_{1\nu }k_{2\mu
}+k_{1\mu }k_{2\nu }+2k_{1\nu }k_{1\mu }\right) \left[ I_{\log
}\left( M^{2}\right) \right]
\notag \\
&&+i(4\pi )^{-2}\left[ \left( k_{1}-k_{2}\right) _{\mu }\left(
k_{1}-k_{2}\right) _{\nu }-g_{\mu \nu }\left( k_{1}-k_{2}\right)
^{2}\right]
\notag \\
&&\times \left[ \frac{1}{4}Z_{0}\left( \left( k_{1}-k_{2}\right)
^{2};M^{2}\right) -Z_{2}\left( \left( k_{1}-k_{2}\right)
^{2};M^{2}\right) \right]  \notag \\
&&-i(4\pi )^{-2}\left( k_{1}+k_{2}\right) _{\mu }\left(
k_{1}+k_{2}\right) _{\nu }\left[ \frac{1}{4}Z_{0}\left( \left(
k_{1}-k_{2}\right) ^{2};M^{2}\right) \right] .
\end{eqnarray}
In these, the functions $Z_0\left( q^{2};M^{2}\right)$ and
$Z_2\left( q^{2};M^{2}\right)$ are finite and can be written
generically as
\begin{equation}
Z_{k}\left( q^{2};M^{2}\right) =\int_{0}^{1} dz \, z^{k} \log
\left[ \frac{ q^{2}z(1-z) - M^{2}}{-M^{2}}\right],
\label{definition-Z0}
\end{equation}
and $\square_{\alpha \beta \mu \nu }$, $\Delta _{\mu \nu}$,
$\nabla_{\mu\nu}$, $I_{log}$ and $I_{quad}$ are
momentum-independent divergent quantities given by
\begin{eqnarray}
\square_{\alpha \beta \mu \nu } &=&\int_{\Lambda
}\frac{d^{4}k}{\left( 2\pi \right) ^{4}}\frac{24k_{\mu }k_{\nu
}k_{\alpha }k_{\beta }}{\left(
k^{2}-M^{2}\right) ^{4}}-g_{\alpha \beta }\int_{\Lambda }\frac{d^{4}k}{
\left( 2\pi \right) ^{4}}\frac{4k_{\mu }k_{\nu }}{\left(
k^{2}-M^{2}\right)
^{3}}  \notag \\
&&-g_{\alpha \nu }\int_{\Lambda }\frac{d^{4}k}{\left( 2\pi \right) ^{4}}
\frac{4k_{\beta }k_{\mu }}{\left( k^{2}-M^{2}\right)
^{3}}-g_{\alpha \mu }\int_{\Lambda }\frac{d^{4}k}{\left( 2\pi
\right) ^{4}}\frac{4k_{\beta
}k_{\nu }}{\left( k^{2}-M^{2}\right) ^{3}}, \\
\Delta _{\mu \nu } &=&\int_{\Lambda }\frac{d^{4}k}{\left( 2\pi \right) ^{4}}
\frac{4k_{\mu }k_{\nu }}{\left( k^{2}-M^{2}\right) ^{3}}-\int_{\Lambda }%
\frac{d^{4}k}{\left( 2\pi \right) ^{4}}\frac{g_{\mu \nu }}{\left(
k^{2}-M^{2}\right) ^{2}}, \\
\nabla _{\mu \nu } &=&\int_{\Lambda }\frac{d^{4}k}{\left( 2\pi \right) ^{4}}
\frac{2k_{\nu }k_{\mu }}{\left( k^{2}-M^{2}\right) ^{2}}-\int_{\Lambda }%
\frac{d^{4}k}{\left( 2\pi \right) ^{4}}\frac{g_{\mu \nu }}{\left(
k^{2}-M^{2}\right) }, \\
I_{log}\left( M^{2}\right) &=&\int_{\Lambda }\frac{d^{4}k}{\left(
2\pi
\right) ^{4}}\frac{1}{\left( k^{2}-M^{2}\right) ^{2}}, \\
I_{quad}\left( M^{2}\right) &=&\int_{\Lambda }\frac{d^{4}k}{\left(
2\pi \right) ^{4}}\frac{1}{\left( k^{2}-M^{2}\right) } .
\end{eqnarray}

Since the one-point functions are purely divergent, they can be
expressed entirely in terms of (a subset of) the above divergent
quantities as:
\begin{eqnarray}
T^{S}\left( k_{1}\right) &=& 4M \left\{ \left[
I_{quad}(M^{2})\right] +k_{1}^{\alpha }k_{1}^{\beta }\left[ \Delta
_{\beta \alpha }\right] \right\},  \label{T-S2} \\
T_{\mu }^{V}\left( k_{1}\right) &=&4 \Bigl\{ -k_{1}^{\beta }\left[
\nabla_{\beta \mu }\right] -\frac{1}{3}k_{1}^{\beta }k_{1}^{\alpha }k_{1}^{\nu }
\left[\square _{\alpha \beta \mu \nu }\right] \notag \\
&&+ \frac{1}{3}k_{1}^{2}k_{1}^{\nu }\left[ \Delta_{\mu \nu}\right]
+ \frac{2}{3}k_{1\mu }k_{1}^{\alpha }k_{1}^{\beta } \left[
\Delta_{\alpha \beta} \right] \Bigr\} . \label{T-V2}
\end{eqnarray}
The two-point functions contain finite and divergent parts, and
can be written as:
\begin{eqnarray}
T^{SS}\left( k_{1},k_{2}\right) &=& 4 \Biggl\{ \left[
I_{quad}(M^{2})\right] + \frac{1}{2}\left[
4M^{2}-(k_{1}-k_{2})^{2}\right] [I_{log}(M^{2})]
\notag \\
&&- i (4\pi )^{-2}\left[ 4M^{2}-\left( k_{1}-k_{2}\right)^{2}
\right] \left[ \frac{1}{2}Z_{0}\left( \left( k_{1}-k_{2}\right)
^{2},M^{2}\right) \right] \Biggr\}  \notag \\
&&+ (k_{1}-k_{2})^{\alpha }(k_{1}-k_{2})^{\beta }\left[ \Delta
_{\alpha\beta }\right]  \notag \\
&&+ (k_{1}+k_{2})^{\alpha }(k_{1}+k_{2})^{\beta }\left[ \Delta
_{\alpha \beta }\right] ,  \label{T-SS2}
\end{eqnarray}
\begin{eqnarray}
T^{PP}\left( k_{1},k_{2}\right) &=& 4\Biggl\{ -\left[
I_{quad}(M^{2})\right] + \frac{1}{2}\left( k_{1}-k_{2}\right)^{2}
[I_{log}(M^{2})] \notag \\
&&- i(4\pi )^{-2}\left( k_{1}-k_{2}\right)^{2}\left[ \frac{1}{2}
Z_{0}\left( \left( k_{1}-k_{2}\right) ^{2},M^{2}\right) \right]
\Biggr\} \notag \\
&&- (k_{1}-k_{2})^{\alpha }(k_{1}-k_{2})^{\beta } \left[
\Delta_{\alpha \beta }\right]  \notag \\
&&- (k_{1}+k_{2})^{\alpha }(k_{1}+k_{2})^{\beta }\left[\Delta
_{\alpha \beta }\right] ,  \label{T-PP2}
\end{eqnarray}
\begin{eqnarray}
T_{\mu }^{PA}\left( k_{1},k_{2}\right) &=& 4M (k_{1}-k_{2})_{\mu
}\left\{ [I_{log}(M^{2})]-i(4\pi )^{-2}\left[ Z_{0}\left( \left(
k_{1}-k_{2}\right)
^{2},M^{2}\right) \right] \right\} ,  \label{T-AP2} \\
T_{\mu }^{VS}\left( k_{1},k_{2}\right) &=&-4M(k_{1}+k_{2})^{\xi
}[\Delta_{\xi \mu }],  \label{T-VS2} \\
T_{\mu \nu }^{AV}\left( k_{1},k_{2}\right) &=&-2i\varepsilon _{\mu \nu
\alpha \beta }(k_{2}-k_{1})^{\beta }(k_{1}+k_{2})^{\xi }\left[ \Delta _{\xi
}^{\alpha }\right] ,  \label{T-AV2}
\end{eqnarray}
\begin{eqnarray}
T_{\mu \nu }^{VV}\left( k_{1},k_{2}\right) &=&\frac{4}{3}[%
(k_{1}-k_{2})^{2}g_{\mu \nu }-(k_{1}-k_{2})_{\mu }(k_{1}-k_{2})_{\nu }
]\left\{ \left[ I_{log}(M^{2})\right] \right.  \notag \\
&&\left. -i(4\pi )^{-2}\left[ \frac{1}{3}+\frac{(2M^{2}+(k_{1}-k_{2})^{2})}{
(k_{1}-k_{2})^{2}}\left[ Z_{0}\left( \left( k_{1}-k_{2}\right)
^{2},M^{2}\right) \right] \right] \right\} + A_{\mu \nu },
\label{T-VV2}
\end{eqnarray}
\begin{eqnarray}
T_{\mu \nu }^{AA}\left( k_{1},k_{2}\right) &=&\frac{4}{3}[%
(k_{1}-k_{2})^{2}g_{\mu \nu }-(k_{1}-k_{2})_{\mu }(k_{1}-k_{2})_{\nu }
]\left\{ \left[ I_{log}(M^{2})\right] \right.  \notag \\
&&\left. -i(4\pi )^{-2}\left[ \frac{1}{3}+\frac{(2M^{2}+(k_{1}-k_{2})^{2})}{
(k_{1}-k_{2})^{2}}\left[ Z_{0}\left( \left( k_{1}-k_{2}\right)
^{2},M^{2}\right) \right] \right] \right\}  \notag \\
&&-8M^{2}g_{\mu \nu }\left\{ [I_{log}(M^{2})]-i(4\pi )^{-2}\left[
Z_{0}\left( \left( k_{1}-k_{2}\right) ^{2},M^{2}\right) \right]
\right\} + A_{\mu \nu }.  \label{T-AA2}
\end{eqnarray}
\begin{eqnarray}
T_{\mu \nu }^{ST}\left( k_{1},k_{2}\right) &=&-2\left\{ \left(
k_{1}-k_{2}\right) _{\mu }\left( k_{1}+k_{2}\right) ^{\xi }\left[ \Delta
_{\xi \nu }\right] -\left( k_{1}-k_{2}\right) _{\nu }\left(
k_{1}+k_{2}\right) ^{\xi }\left[ \Delta _{\xi \mu }\right] \right\} ,
\label{T-ST2} \\
T_{\mu \nu }^{PT}\left( k_{1},k_{2}\right) &=& -2i\varepsilon
_{\mu \nu \alpha \beta }\left( k_{1}-k_{2}\right) ^{\alpha }\left(
k_{1}+k_{2}\right) ^{\xi }\left[ \Delta _{\xi }^{\beta }\right] ,
\label{T-PT2}
\end{eqnarray}
\begin{eqnarray}
T_{\alpha \mu \nu }^{AT}\left( k_{1},k_{2}\right)
&=&4iM\varepsilon _{\mu \nu \alpha \beta }(k_{1}+k_{2})^{\xi
}\left[ \Delta _{\xi }^{\beta }\right] ,
\label{T-AT2} \\
T_{\alpha \mu \nu }^{VT}\left( k_{1},k_{2}\right) &=& 4M\left(
g_{\alpha \nu } g_{\mu \lambda } - g_{\alpha \mu } g_{\nu \lambda
}\right) \left(k_{1}-k_{2}\right)^{\lambda }  \notag \\
&&\times \left\{ \lbrack I_{log}(M^{2})]-i(4\pi )^{-2}\left[
Z_{0}\left( \left( k_{1}-k_{2}\right)^{2},M^{2}\right) \right]
\right\} ,  \label{T-VT2}
\end{eqnarray}
\begin{eqnarray}
T_{\alpha \beta \mu \nu }^{TT}\left( k_{1},k_{2}\right) &=&\left(
g_{\alpha \mu } g_{\beta \xi } g_{\nu \lambda } - g_{\alpha \nu }
g_{\beta \xi }g_{\mu \lambda } + g_{\beta \nu }g_{\alpha \xi }
g_{\mu \lambda }-g_{\beta \mu} g_{\alpha \xi } g_{\nu \lambda }\right)  \notag \\
&&\times \frac{4}{3}[\left( k_{1}-k_{2}\right)^{2} g^{\xi \lambda} -
\left(k_{1}-k_{2}\right) ^{\xi }\left( k_{1}-k_{2}\right) ^{\lambda }]  \notag \\
&&\times \left\{ I_{log}(m^{2})-i(4\pi )^{-2}\left[
\frac{1}{3}+\frac{2M^{2} + \left( k_{1}-k_{2}\right) ^{2}}{\left(
k_{1}-k_{2}\right) ^{2}} \left[ Z_{0}\left( M^{2},\left(
k_{1}-k_{2}\right) ^{2}\right) \right] \right]
\right\}  \notag \\
&&+4\left( g_{\alpha \nu }g_{\beta \mu } - g_{\alpha \mu }
g_{\beta \nu } \right) \left\{ -\left[ I_{quad}(M^{2})\right]
+\frac{1}{2}\left[4M^{2}+(k_{1}-k_{2})^{2}\right] [I_{log}(M^{2})]\right. \notag \\
&&-\frac{1}{2}\left[ 4M^{2}+\left( k_{1}-k_{2}\right) ^{2}\right]
i(4\pi )^{-2}\left[ Z_{0}\left( \left( k_{1}-k_{2}\right)
^{2},M^{2}\right) \right]
\notag \\
&&\left. -\frac{1}{4}(k_{1}-k_{2})^{\lambda }(k_{1}-k_{2})^{\xi }\left[
\Delta _{\lambda \xi }\right] -\frac{1}{4}(k_{1}+k_{2})^{\lambda
}(k_{1}+k_{2})^{\xi }\left[ \Delta _{\lambda \xi }\right] \right\}  \notag \\
&&+g_{\alpha \mu }A_{\beta \nu }-g_{\alpha \nu }A_{\beta \mu
}+g_{\beta \nu }A_{\alpha \mu }-g_{\beta \mu }A_{\alpha \nu } .
\label{T-TT2}
\end{eqnarray}
In the above, we have defined $A_{\mu \nu }$ as
\begin{eqnarray}
A_{\mu \nu } &=&4[\nabla _{\mu \nu }]+(k_{1}-k_{2})^{\alpha
}(k_{1}-k_{2})^{\beta }  \notag \\
&&\times \left[ \frac{1}{3}\square _{\alpha \beta \mu \nu }+\frac{1}{3}%
g_{\alpha \nu }\Delta _{\mu \beta }+g_{\alpha \mu }\Delta _{\beta \nu
}-g_{\mu \nu }\Delta _{\alpha \beta }-\frac{2}{3}g_{\alpha \beta }\Delta
_{\mu \nu }\right]  \notag \\
&&+\left[ (k_{1}-k_{2})^{\alpha }(k_{1}+k_{2})^{\beta
}-(k_{1}+k_{2})^{\alpha }(k_{1}-k_{2})^{\beta }\right]  \notag \\
&&\times \left[ \frac{1}{3}\square _{\alpha \beta \mu \nu }+\frac{1}{3}%
g_{\nu \alpha }\Delta _{\mu \beta }+\frac{1}{3}g_{\alpha \mu }\Delta _{\beta
\nu }\right]  \notag \\
&&+(k_{1}+k_{2})^{\alpha }(k_{1}+k_{2})^{\beta }\left[ \square
_{\alpha \beta \mu \nu }-g_{\mu \beta }\Delta _{\nu \alpha
}-g_{\alpha \mu }\Delta _{\beta \nu }-3g_{\mu \nu }\Delta _{\alpha
\beta }\right] . \label{Amunu}
\end{eqnarray}
The remaining one- and two-point functions not considered above
are identically zero due to vanishing Dirac traces.

At this point it is important to note the generality of the
method. No momentum shifts were done and no single divergent
integral was calculated. For this reason, the results obtained can
be used with any preferred regularization method. Of course,
different regularization schemes can lead to ambiguities and
violation of symmetry relations between Green's functions, as we
shall discuss in the next section.

\section{Ambiguities and the consistency constraints}
\label{sec:ambiguities}

In the preceding section we have completed the evaluation of all
fermionic one- and two-point amplitudes. Before any further steps
in evaluating these amplitudes, it is important to notice that
there are ambiguities associated with the choices for the momentum
routing in the internal lines of loops. In the calculations
performed in the previous section, we have not made any momentum
shifts in intermediary steps and have left the labels of the
internal momenta $k_1$ and $k_2$ completely arbitrary and
unspecified. However, due to momentum conservation a physical
amplitude can depend only on the difference $q \equiv
k_{1}-k_{2}$, and no dependence on the sum $Q \equiv k_{1}+k_{2}$
can be present in the amplitude. There can be no dependence on $Q$
because this combination is ambiguous: two different choices of
$k_1$ and $k_2$ can give the same $q$, but they will always give
different $Q$'s. In view of this, one could imagine that it is
impossible to make any predictions because the amplitudes contain
dependencies on arbitrary contributions which are functions of
$Q$. In general, depending on the regularization scheme used, this
is actually true. In the literature associated with the NJL model,
it is usual to use a particular routing, the symmetric combination
$k_1 = q/2$ and $k_2 = -q/2$, meaning that $Q = 0$ and the
amplitudes are automatically free of ambiguities. However,
restriction to one particular choice, besides breaking homogeneity
of space-time, also leads to difficulties with amplitudes containing
more than two propagators where there are in principle more than
two arbitrary momenta and choices like the above would not be
allowed because in general would lead to violations of momentum
conservation.

Our aim now is to show that what we have done so far does not lead
to violations of the relations among Green's functions because of
arbitrary choices of the momenta $k_1$ and $k_2$. As shown in
Section~\ref{sec:consistency}, such relations are very general,
but due to the ultraviolet divergences they can very easily be
violated when not being careful with the explicit evaluation of
the integrals within a particular regularization scheme.

It is a simple task to identify the ambiguous terms, proportional
to $Q$. In the one-point functions they are given by (note that
for these $Q = k_1$):
\begin{eqnarray}
\left[ T^{S}\left( k_{1}\right) \right] _{ambi} &=& 4 M
k_{1}^{\beta} k_{1}^{\alpha } \left[ \Delta _{\beta \alpha }\right] , \\
\left[ T_{\mu }^{V}\left( k_{1}\right) \right]_{ambi} &=& 4
\Bigl\{ -k_{1}^{\beta }\left[ \nabla _{\beta \mu }\right] -
\frac{1}{3}k_{1}^{\beta }k_{1}^{\alpha }k_{1}^{\nu }
\left[ \square_{\alpha \beta \mu \nu }\right]   \notag \\
&&+ \frac{1}{3}k_{1}^{2}k_{1}^{\nu }\left[ \Delta _{\mu \nu}
\right] + \frac{2}{3}k_{1\mu }k_{1}^{\alpha }k_{1}^{\beta }\left[
\Delta _{\alpha \beta }\right] \Bigr\} .
\end{eqnarray}
In the two-point functions, the ambiguous terms are given by:
\begin{eqnarray}
\left[ T^{SS}\left( k_{1},k_{2}\right) \right] _{ambi} &=&Q^{\alpha
}Q^{\beta }\left[ \Delta _{\alpha \beta }\right] , \\
\left[ T^{PP}\left( k_{1},k_{2}\right) \right] _{ambi} &=&-\left[
T^{SS}\left( k_{1},k_{2}\right) \right] _{ambi}, \\
\left[ T_{\mu }^{PA}\left( k_{1},k_{2}\right) \right] _{ambi} &=&0, \\
\left[ T_{\mu }^{VS}\left( k_{1},k_{2}\right) \right] _{ambi}
&=&-4MQ^{\xi }[\Delta _{\xi \mu }],
\end{eqnarray}
\begin{eqnarray}
\left[ T_{\mu \nu }^{AV}\left( k_{1},k_{2}\right) \right] _{ambi}
&=&2i\varepsilon _{\mu \nu \alpha \beta }q^{\beta }Q^{\xi }\left[ \Delta
_{\xi }^{\alpha }\right] , \\
\left[ T_{\mu \nu }^{VV}\left( k_{1},k_{2}\right) \right] _{ambi} &=&\left[
q^{\alpha }Q^{\beta }-Q^{\alpha }q^{\beta }\right] \left[ \frac{1}{3}\square
_{\alpha \beta \mu \nu }+\frac{1}{3}g_{\nu \alpha }\Delta _{\mu \beta }+%
\frac{1}{3}g_{\alpha \mu }\Delta _{\beta \nu }\right]  \notag \\
&&+ Q^{\alpha }Q^{\beta }\left[ \square _{\alpha \beta \mu \nu } -
g_{\mu \beta }\Delta _{\nu \alpha }-g_{\alpha \mu }\Delta _{\beta
\nu }-3g_{\mu \nu
}\Delta _{\alpha \beta }\right], \\
\left[ T_{\mu \nu }^{AA}\left( k_{1},k_{2}\right) \right] _{ambi}
&=&\left[ T_{\mu \nu }^{VV}\left( k_{1},k_{2}\right) \right]
_{ambi},
\end{eqnarray}
\begin{eqnarray}
\left[ T_{\mu \nu }^{ST}\left( k_{1},k_{2}\right) \right] _{ambi}
&=&-2\left\{ q_{\mu }Q^{\xi }\left[ \Delta _{\xi \nu }\right] -q_{\nu
}Q^{\xi }\left[ \Delta _{\xi \mu }\right] \right\} , \\
\left[ T_{\mu \nu }^{PT}\left( k_{1},k_{2}\right) \right] _{ambi}
&=&-i2\varepsilon _{\mu \nu \alpha \beta }q^{\alpha }Q^{\xi }\left[ \Delta
_{\xi }^{\beta }\right] , \\
\left[ T_{\alpha \mu \nu }^{AT}\left( k_{1},k_{2}\right) \right]
_{ambi} &=&4iM\varepsilon _{\mu \nu \alpha \beta }Q^{\xi }\left[
\Delta _{\xi
}^{\beta }\right] , \\
\left[ T_{\alpha \mu \nu }^{VT}\left( k_{1},k_{2}\right) \right] _{ambi}
&=&0,
\end{eqnarray}
\begin{eqnarray}
\left[ T_{\alpha \beta \mu \nu }^{TT}\left( k_{1},k_{2}\right) \right]
_{ambi} &=&-\frac{1}{4}Q^{\lambda }Q^{\xi }\left[ \Delta _{\lambda \xi }%
\right]  \notag \\
&&+ g_{\alpha \mu }\left( T_{\beta \nu }^{VV}\right)
_{ambi}-g_{\alpha \nu
}\left( T_{\beta \mu }^{VV}\right) _{ambi}  \notag \\
&&+ g_{\beta \nu }\left( T_{\alpha \mu }^{VV}\right)
_{ambi}-g_{\beta \mu }\left( T_{\alpha \nu }^{VV}\right) _{ambi}
\end{eqnarray}

Note that all ambiguous terms appear as coefficients of the
divergent quantities $\ \nabla _{\beta \mu },\;\square _{\alpha
\beta \mu \nu }$ and $\triangle _{\alpha \beta }$. These
quantities also play an important role in the analysis of the
algebraic relations involving different Green's functions, the
consistency constrains obtained in Section~\ref{sec:consistency}.
Those constraints were obtained by making the only assumption of
the validity of the linearity of the integration operation.
Although this assumption seems very reasonable, the mathematical
indefiniteness due to the divergences turns the preservation of
such relations a non trivial supposition.

Let us now show that despite the amplitudes themselves contain
ambiguous terms, the general relations involving different Green's
functions obtained in Section~\ref{sec:consistency} are not
violated by the manipulations done so far. We start with the
relations in Eqs.~(\ref{rel1-T-VV}) and (\ref{rel-T-VV}) involving
the Green's functions $T_{\mu\nu}^{VV} \left( k_{1},k_{2}\right)$
and $T_{\nu }^{V}\left( k_{1}\right)$. Contracting $T_{\mu \nu
}^{VV}\left( k_{1},k_{2}\right)$ given in Eq.~(\ref{T-VV2}) with
the external momentum $q$ leads to
\begin{equation}
q^{\mu }T_{\mu \nu }^{VV}\left( k_{1},k_{2}\right) =q^{\mu }A_{\mu
\nu }. \label{rel-Amunu}
\end{equation}
From Eq.~(\ref{rel-T-VV}), one sees that one must identify on the
r.h.s. of Eq.~(\ref{rel-Amunu}) the difference between two vector
one-point functions corresponding to the internal propagator
carrying arbitrary momenta $k_{1}$ and $k_{2}$. For this purpose
we take Eq.~(\ref{Amunu}) for $A_{\mu\nu}$ and contract it with
$q^\mu$
\begin{equation}
q^{\mu }A_{\mu \nu }=T_{\nu }^{V}(k_{2})-T_{\nu }^{V}(k_{1}),
\end{equation}
and also with $q^\nu$
\begin{equation}
q^{\nu }A_{\mu \nu }=T_{\mu }^{V}(k_{2})-T_{\mu }^{V}(k_{1}),
\end{equation}
where we used Eq.~(\ref{T-V2}) for identifying $T_{\nu }^{V}
\left(k_{1}\right)$. These results imply that the algebraic
relations of Eqs.~(\ref{rel1-T-VV}) and (\ref{rel-T-VV}) involving
$T_{\nu }^{V}\left( k_{1}\right)$ and $T_{\mu \nu }^{VV}\left(
k_{1},k_{2}\right)$ obtained by the formal manipulations in
Section~\ref{sec:consistency} are preserved by the explicit and
independent evaluations of $T_{\mu \nu}^{VV} \left(
k_{1},k_{2}\right) $ and $T_{\mu }^{V}\left( k_{1}\right)$ in
Section~\ref{sec:isolating}. Therefore, the relations are
preserved in spite of the fact that both amplitudes $T_{\nu
}^{V}\left( k_{1}\right)$ and $T_{\mu \nu }^{VV}\left(
k_{1},k_{2}\right)$ have ambiguous pieces, which are proportional
to the divergent quantities $\nabla _{\beta \mu },\;\square
_{\alpha \beta \mu \nu }$ and $\triangle _{\alpha \beta }$.

Next we consider the relation given in Eq.~(\ref{rel-T-VS})
involving the amplitudes $T^{VS}_\mu$ and $T^S$. Using the
explicit expression for $T_{\mu }^{VS}\left( k_{1},k_{2}\right)$
given in Eq.~(\ref{T-VS2}), one obtains
\begin{equation}
q^{\mu }T_{\mu }^{VS}\left( k_{1},k_{2}\right) =-4M\left(
k_{1}^{\mu }k_{1}^{\xi }+k_{1}^{\mu }k_{2}^{\xi }-k_{2}^{\mu
}k_{1}^{\xi }-k_{2}^{\mu }k_{2}^{\xi }\right) [\Delta _{\xi \mu
}].
\end{equation}
Due to the obvious symmetry under interchange of the Lorentz
indexes $\xi$ and $\mu$ in $\Delta _{\xi \mu }$, and by comparing
with the explicit result for $T^S$ in Eq.~(\ref{T-S2}), it is very
simple to show that the identity of Eq.~(\ref{rel-T-VS}) is also
preserved.

Now we consider the identity given in Eq.~(\ref{rel-T-AA}),
involving the amplitudes $T_{\mu \nu }^{AA}$, $T^{AP}_\mu$ and
$T^V_\mu$. Contracting $T_{\mu \nu }^{AA}$ given in
Eq.~(\ref{T-AA2}) with $q^\mu$, we get
\begin{eqnarray}
q^{\mu }T_{\mu \nu }^{AA}\left( k_1, k_{2}\right) &=& 2M \, \left
\{ - 4M \, q_{\nu} \, \left[ I_{log}\left( M^{2}\right) -i(4\pi
)^{-2}Z_{0}\left( q^{2};M^{2}\right) \right] \right\} + q^{\mu}
A_{\mu \nu }.
\end{eqnarray}
It is easy to verify that using Eqs.~(\ref{T-AP2}) and
(\ref{T-V2}), the identity in Eq.~(\ref{rel-T-AA}) is preserved.

We proceed examining the relation given in Eq.~(\ref{rel-T-AP})
using Eq.~(\ref{T-AP2}) for the amplitude $T_{\mu}^{AP} \left(
k_{1}, k_{2}\right)$. Adding and subtracting scalar one-point
functions carrying momenta $k_{1}\ $and $k_{2}$, one can easily
check that this expression for $T^{PP}\left( k_{1},k_{2}\right)$
naturally leads to preservation of the relation in
Eq.~(\ref{rel-T-AP}). Also, the relations involving the $AV$
two-point function in Eqs.~(\ref{rel-T-AV1}) and (\ref{rel-T-AV2})
are immediate: the Green's functions $T^{PV}_\mu$ and and
$T^A_\mu$ on the r.h.s. of Eqs.~(\ref{rel-T-AV1}) and
(\ref{rel-T-AV2}) are identically zero due to properties the Dirac
traces, and on the l.h.s. one also obtains a zero because the
contractions of the explicit expression for $T_{\mu \nu
}^{AV}\left( k_{1},k_{2}\right)$ in Eq.~(\ref {T-AV2}) with $(k_1
- k_2)^{\mu}$ or $(k_1 - k_2)^{\mu}$ are zero due to the
antisymmetry of $\varepsilon _{\mu \nu \lambda \xi }$.

Next, we turn our attention to the tensorial amplitudes. We start
with the relations of Eqs.~(\ref{rel-T-VT1}) and
(\ref{rel-T-VT2}). Initially we note that the amplitude $T_{\alpha
\mu \nu }^{VT}\left( k_{1},k_{2}\right)$ given in
Eq.~(\ref{T-VT2}) can be reorganized as
\begin{eqnarray}
T_{\alpha \mu \nu }^{VT}\left( k_{1},k_{2}\right) &=&\frac{1}{2M}
\, q_{\mu } \, \left[ T_{\alpha \nu }^{VV}\left(
k_{1},k_{2}\right) -
T_{\alpha\nu}^{AA} \left( k_{1},k_{2}\right) \right]  \notag \\
&-& \frac{1}{2M}\, q_{\nu } \, \left[ T_{\alpha \mu }^{VV}\left(
k_{1},k_{2}\right) -T_{\alpha \mu }^{AA}\left( k_{1},k_{2}\right)
\right] ,
\end{eqnarray}
where we have used Eqs.~(\ref{T-VV2}) and\ (\ref{T-AA2}). Written
in this form, it is now trivial to see that Eqs.~(\ref{rel-T-VT1})
and (\ref{rel-T-VT2}) are satisfied. Next we consider the relation
in Eq.~(\ref{rel-T-TT}). The $TT$ two-point function, given by
Eq.~(\ref{T-TT2}), after some algebraic effort can be put in the
form
\begin{eqnarray}
T_{\alpha \beta \mu \nu }^{TT}\left( k_{1},k_{2}\right)
&=&g_{\alpha \mu }T_{\beta \nu }^{AA}\left( k_{1},k_{2}\right)
-g_{\alpha \nu} T_{\beta \mu}^{AA} \left( k_{1},k_{2}\right)  \notag \\
&&+g_{\beta \nu }T_{\alpha \mu }^{AA} \left( k_{1},k_{2}\right) -
g_{\beta \mu
}T_{\alpha \nu }^{AA}\left( k_{1},k_{2}\right)  \notag \\
&&+\left( g_{\alpha \mu }g_{\beta \nu } - g_{\alpha \nu }
g_{\beta\mu }\right) T^{SS} \left( k_{1},k_{2}\right) ,
\end{eqnarray}
where the Eqs.(\ref{T-AA2}) and (\ref{T-SS2}) have been used. It
is now evident that the contraction with the external momentum
leads to Eq.~(\ref{rel-T-TT}). Now we consider the constraint
given in Eq.~(\ref{rel-T-TS}). First we note that
Eq.~(\ref{T-ST2}) for the $ST$ amplitude, by using
Eqs.~(\ref{T-VS2}) and (\ref{ident1}), can be rewritten as
\begin{eqnarray} T_{\mu \nu }^{ST}\left(
k_{1},k_{2}\right) &=& \frac{1}{2M}\left[ q_{\mu }T_{\nu
}^{SV}\left( k_{1},k_{2}\right) - q_{\nu }T_{\mu }^{SV}\left(
k_{1},k_{2}\right) \right] ,  \notag \\
&=&-\frac{i}{2}\varepsilon _{\mu \nu \lambda \xi }\left( T^{AV}\right)
^{\lambda \xi }\left( k_{1},k_{2}\right) .
\end{eqnarray}
Therefore, contraction with the external momentum
\begin{eqnarray}
q^{\mu }T_{\mu \nu }^{TS}\left( k_{1},k_{2}\right) &=&
\frac{1}{2M}\left\{ q^{2}T_{\nu }^{SV}\left( k_{1},k_{2}\right)
-q_{\nu }\left[ T^{S}\left(
k_{2}\right) -T^{S}\left( k_{1}\right) \right] \right\} ,  \notag \\
&=&-\frac{i}{2}\varepsilon _{\mu \nu \lambda \xi }q^{\mu }\left(
T^{AV}\right) ^{\lambda \xi }\left( k_{1},k_{2}\right) ,
\end{eqnarray}
leads immediately to Eq.~(\ref{rel-T-TS}). Now, noting that by
Eqs.~(\ref{T-AV2}) and (\ref{T-PT2}) the $AV$ Green's function is
identical to the $TP$ function, we have that
\begin{equation}
q^{\mu }T_{\mu \nu }^{TP}\left( k_{1},k_{2}\right) =-2MT_{\nu
}^{VP}\left( k_{1},k_{2}\right) ,
\end{equation}
as it should to satisfy Eq.~(\ref{rel-T-TP}). Also, comparing
Eqs.~(\ref{T-VS2}) and (\ref{T-AT2}), one sees that
\begin{equation}
T_{\alpha \mu \nu }^{AT}\left( k_{1},k_{2}\right) =-i\varepsilon
_{\mu \nu \alpha \beta }\left( T^{SV}\right) ^{\beta }\left(
k_{1},k_{2}\right),
\end{equation}
and then
\begin{eqnarray}
q^{\alpha }T_{\mu \nu \alpha }^{TA}\left( k_{1},k_{2}\right) &=&
2MT_{\mu \nu}^{TP}\left( k_{1},k_{2}\right) , \\
q^{\mu }T_{\mu \nu \alpha }^{TA}\left( k_{1},k_{2}\right) &=&
-2MT_{\nu\alpha }^{VA} \left( k_{1},k_{2}\right) = -i\varepsilon
_{\nu \alpha \lambda \xi }q^{\lambda }\left( T^{SV}\right) ^{\xi
}\left( k_{1},k_{2}\right) .
\end{eqnarray}
Therefore, the remaining relations, given by
Eqs.~(\ref{rel-T-TA1}) and (\ref{rel-T-TA2}), are also satisfied.

This completes the verification of the consistency of the
manipulations performed at the one loop level. It is important to
emphasize that all the relations among Green's functions which can
be stated at the level of integrands are preserved, in spite of
the presence of ambiguous terms. The arbitrariness concerning the
choice of the regularization is also preserved since only very
general mathematical properties have been assumed.

The crucial point now is that the preservation of the relations
among Green's functions is not the only requirement one should ask
for a consistent regularization scheme since, as we have seen, the
amplitudes themselves contain ambiguities that are functions of
the ambiguous combination of momenta $Q = k_1 + k_2$. Therefore,
one must have a scheme that ensures elimination of the ambiguous
terms from the amplitudes themselves, not only in the relations
involving two or more of them. In~the next section we discuss the
constraints imposed by symmetry relations (Furry's theorem and
Ward identities) on special amplitudes and they will provide
guidance for dealing with the ambiguous terms.

\section{Ambiguities and Preservation of Symmetries}
\label{sec:symmetries}

In the process of constructing a consistent interpretation for the
divergent one-loop amplitudes the preservation of symmetries plays
a central role. In principle, there is no a priori reason for
expecting that space-time symmetries will be automatically
manifest in divergent amplitudes. However, it seems nevertheless
reasonable to expect that one should be able to identify general
properties that the divergent quantities must satisfy in order to
guarantee the preservation of such fundamental symmetries. In this
sense, the regularization method itself is not the most important
ingredient, what really matters are the requirements that
quantities like $\ \nabla _{\beta \mu },\;\square _{\alpha \beta
\mu \nu }$ and $\triangle _{\alpha \beta }$ must obey to preserve
the symmetry relations. Having
this in mind let us now consider the symmetry properties pertinent
to the one- and two-point amplitudes we are discussing. We shall
refer to the Ward identities and other general constraints imposed
by Furry's theorem on these amplitudes.

We start considering the simplest amplitude that carries one
Lorentz vector index, the amplitude $T_{\mu }^{V}\left(
k_{1}\right)$. On general symmetry grounds, Furry's theorem states
that this amplitude must be zero. So, from Eq.~(\ref{T-V2}),
Furry's theorem requires that
\begin{equation}
T^V_\mu = - k_{1}^{\beta }\nabla _{\beta \mu
}-\frac{1}{3}k_{1}^{\beta} k_{1}^{\alpha}k_{1}^{\nu } \left[
\square_{\alpha \beta \mu \nu } \right] +
\frac{1}{3}k_{1}^{2}k_{1}^{\nu }\left[ \triangle _{\nu \mu
}\right] + \frac{2}{3}k_{1\mu }k_{1}^{\alpha }k_{1}^{\beta }
\left[ \triangle _{\alpha \beta }\right] = 0. \label{T-V=0}
\end{equation}
There are two different ways to satisfy this requirement. The
first one is the choice $k_{1}=0$. But, is it always possible to
make this choice? Thinking on $T^V_\mu$ in isolation, the answer
to this question is affirmative, since $k_{1}$ is arbitrary.
However, $T^V_\mu$ is not the only amplitude in the theory and so
one must ask the question if this choice is not invalidating other
symmetry relations. For example, the relation given in
Eq.~(\ref{rel-T-VV}) relates this amplitude to $ \left(
k_{1}-k_{2}\right)^\mu T^{VV}_{\mu\nu} \left( k_{1},k_{2}\right)$,
which we repeat here for clarity, is given by
\begin{equation}
\left( k_{1}-k_{2}\right) ^{\mu }T_{\mu \nu }^{VV}\left(
k_{1},k_{2}\right) = T_{\nu }^{V}(k_{1})-T_{\nu }^{V}(k_{2}).
\label{TVV-TV}\end{equation}
Vector current conservation demands that $\left(
k_{1}-k_{2}\right)^\mu T^{VV}_{\mu\nu} \left(
k_{1},k_{2}\right)=0$. Therefore, the difference of the two
one-point functions on the r.h.s. of Eq.~(\ref{TVV-TV}) having
dependencies on $ k_{1}$ and $k_{2}$ needs to be zero. Obviously,
the simultaneous choice $k_{1}=0$ and $k_{2}=0$, which would
satisfy both requirements, cannot be made because this would imply
$q = k_1 - k_2 = 0$ always. Therefore, we need another way to
satisfy Eq.~(\ref{T-V=0}). Since the requirement of
Eq.~(\ref{T-V=0}) involves the divergent quantities $\square
_{\alpha \beta \mu \nu }$,  $\nabla _{\mu \nu }$ and $\triangle
_{\mu \nu }$, one could ask for a regularization scheme that leads
to
\begin{equation}
\square _{\alpha \beta \mu \nu }^{reg}=\nabla _{\mu \nu }^{reg} =
\triangle _{\mu \nu }^{reg}=0,  \label{CR}
\end{equation}
where the superscript $reg$ means that the integrals defining
these quantities are regularized.

The same conclusion is reached considering the explicit expression
for $T_{\mu \nu }^{VV}\left( k_{1},k_{2}\right) $ given in
Eq.~(\ref{T-VV2}). Contracting it with the external momentum
$q^\mu = \left( k_{1}-k_{2}\right)^{\mu }$, one obtains:
\begin{eqnarray}
q^\mu T_{\mu \nu }^{VV}\left( k_{1},k_{2}\right) &=&4\left\{
q^{\alpha } \left[ \nabla _{\alpha \nu }\right] +(k_{1}^{\alpha
}k_{1}^{\beta }k_{1}^{\rho }-k_{2}^{\alpha }k_{2}^{\beta
}k_{2}^{\rho })\frac{1}{3}\left[
\square _{\alpha \beta \rho \nu }\right] \right.  \notag \\
&&\;\left. -(k_{1}^{2}k_{1}^{\rho }-k_{2}^{2}k_{2}^{\rho })\frac{1}{3}\left[
\triangle _{\rho \nu }\right] -(k_{1\nu }k_{1}^{\alpha }k_{1}^{\beta
}-k_{2\nu }k_{2}^{\alpha }k_{2}^{\beta })\frac{2}{3}\left[ \triangle
_{\alpha \beta }\right] \right\} .
\end{eqnarray}
Since a conserved vector current should not be obtained by
convenient choices of the arbitrary momenta $k_{1}$ and $k_{2}$,
the conditions of Eq.~(\ref{CR}) seem therefore also necessary
here.

For the same reason that $ T_{\mu }^{V}\left( k_{1}\right)$ must
vanish, other vector two-point functions need vanish identically. 
These are $T_{\mu }^{VS}\left(k_{1},k_{2}\right)$ and $T_{\mu \nu }^{AV}\left(
k_{1},k_{2}\right)$ which, from Eqs.~(\ref{T-VS2}) and
(\ref{T-AV2}), imply in
\begin{eqnarray}
Q^{\xi }\left[ \triangle _{\mu \xi }\right] &=&0, \\
\varepsilon _{\mu \nu \alpha \beta }q^{\beta }Q_{\xi }\left[ \triangle ^{\xi
\alpha }\right] &=&0.
\end{eqnarray}
In principle, for these two specific amplitudes both options, of
choosing $ k_{1}$ and $k_{2}$ in a convenient way or constructing
$\triangle _{\mu \beta }^{reg}=0$, are possible. For example,
considering the contractions of these amplitudes with external
momenta we obtain
\begin{eqnarray}
q^{\mu }T_{\mu }^{VS}\left( k_{1},k_{2}\right) &=& - 4 M q^{\mu}
Q^{\beta }
\left[ \triangle _{\mu \beta }\right] , \\
q^{\mu }T_{\mu \nu }^{AV}\left( k_{1},k_{2}\right) &=&-2\varepsilon _{\mu
\nu \alpha \beta }q^{\mu }q^{\beta }Q_{\xi }\left[ \triangle ^{\xi \alpha }%
\right] , \\
q^{\nu }T_{\mu \nu }^{AV}\left( k_{1},k_{2}\right) &=&-2\varepsilon _{\mu
\nu \alpha \beta }q^{\nu }q^{\beta }Q_{\xi }\left[ \triangle ^{\xi \alpha }%
\right] .
\end{eqnarray}
A conserved vector current for $T_{\mu }^{VS}\left( k_{1},
k_{2}\right)$ can be obtained with the choice $k_{1}=-k_{2}$, or
by taking $\triangle _{\mu \nu }^{reg}=0$. However, both
contractions involving $T_{\mu \nu }^{AV}\left( k_{1},k_{2}\right)
$ vanish identically independently of the two possible choices,
just because the antisymmetric $\varepsilon _{\mu \nu \alpha \beta
}$ is contracted with a symmetric object. The vector current must
be conserved, but the axial-vector current must not. So there is
only one consistent value for $T_{\mu \nu }^{AV}\left(
k_{1},k_{2}\right) $: the identically zero value. Otherwise a
symmetry relation is broken. We can add to this argumentation
another very general aspect that forces us to obtain a zero value
for $T_{\mu \nu }^{AV}\left( k_{1},k_{2}\right) $ (and $T_{\mu
}^{VS}$): unitarity. If the amplitude does not vanish then it
needs to develop an imaginary part at $q^{2}=4M^{2}$ to be
consistent with unitarity (Cutkosky's rules). Clearly, from
Eqs.~(\ref{T-VS2}) and (\ref{T-AV2}) for $T_{\mu }^{VS}\left(
k_{1},k_{2}\right) $ and $T_{\mu \nu }^{AV}\left(
k_{1},k_{2}\right) $, respectively, this cannot happen.

Next we consider $T_{\mu \nu }^{AA}\left( k_{1},k_{2}\right)$.
Using its explicit expression given in Eq.~(\ref{T-AA2}), one can
show that
\begin{eqnarray}
q^{\mu }T_{\mu \nu }^{AA}\left( k_{1},k_{2}\right) &=&4\left\{ -q^{\alpha }
\left[ \nabla _{\alpha \nu }\right] +\left( k_{1}^{\alpha }k_{1}^{\beta
}k_{1}^{\rho }-k_{2}^{\alpha }k_{2}^{\beta }k_{2}^{\rho }\right) \frac{1}{3}%
\left[ \square _{\alpha \beta \rho \nu }\right] \right.  \notag \\
&&\;\left. +\left( k_{1}^{2}k_{1}^{\rho }-k_{2}^{2}k_{2}^{\rho }\right)
\frac{1}{3}\left[ \triangle _{\rho \nu }\right] +\left( k_{1\nu
}k_{1}^{\alpha }k_{1}^{\beta }-k_{2\nu }k_{2}^{\alpha }k_{2}^{\beta }\right)
\frac{2}{3}\left[ \triangle _{\alpha \beta }\right] \right\}  \notag \\
&& - 2 M iT_{\nu }^{PA} \left( k_{1},k_{2}\right) .
\end{eqnarray}
However, the proportionality between the axial-vector and the
pseudoscalar current states that $q^{\mu }T_{\mu \nu}^{AA} \left(
k_{1},k_{2}\right) = - 2 M i T_{\nu }^{PA}\left( k_{1},
k_{2}\right)$. Therefore, one arrives at the same conclusion as
for the amplitude $T_{\mu \nu }^{VV}\left( k_{1},k_{2}\right) $,
that the relations given in Eq.~(\ref{CR}) must be satisfied,
since $T_{\nu }^{PA}\left( k_{1},k_{2}\right) $ is free from
ambiguities -- see Eq.~(\ref{T-AP2}).  Also, the same conclusion
is obtained if one rewrites $q^{\mu }T_{\mu \nu }^{AA}$ in terms
of the amplitudes $T_{\nu }^{V}\left( k_{1}\right)$, as in
Eq.~(\ref{rel-T-AA}).

Considering all amplitudes and their symmetry relations, the same
conditions will emerge: there is no consistent interpretation for
the one-loop divergent amplitudes if the conditions given in
Eq.~(\ref{CR}), that we call {\em consistency relations} (CR), are
not fulfilled. In principle, one could argue that the imposition
of the CR's represents an arbitrary choice which is at the same
level of the choice of a specific regulating distribution to be
used in the integrands of the divergent integrals. However, this
is not true. One should consider the CR's as a fundamental
requirement to be imposed on the one-loop divergent amplitudes in
order to materialize the fundamental space-time symmetries. Any
calculation that violates the CR's has the potential of predicting
unphysical results, since it leads to the destruction of the
foundations of the theory which have generated the amplitudes
themselves. Therefore, the CR's do not represent arbitrary
choices, because there is no option out of these properties
capable to allow a consistent interpretation of the calculations.
Note that the CR's not only remove all the ambiguous terms, which
are always symmetry-violating, but also remove all the symmetry
violating terms, which are not always ambiguous.

To finalize this section, we summarize results by defining what we
denominate the \textquotedblleft consistently regularized
amplitudes \textquotedblright\, (CRA) denoted by $\mathcal{T}^{S},
\mathcal{T}_{\mu }^{VS}, \mathcal{T}^{SS}, \cdots$. These are
respectively the amplitudes $T^{S}, T_{\mu }^{VS}, T^{SS}, \cdots$
obtained previously, with the terms containing the pieces
proportional to the quantities $\square _{\alpha \beta \mu \nu
},\nabla _{\mu \nu },$ and $\triangle _{\mu \nu }$ removed, as
demanded by the CR's given in Eq.~(\ref{CR}). Explicitly, they are
given by:

I) One point functions:
\begin{eqnarray}
\mathcal{T}^{S} &=&4M\left[ I_{quad}(M^{2})\right] ,  \label{T-S} \\
\mathcal{T}_{\mu }^{V} &=&0.  \label{T-V}
\end{eqnarray}

II) Two point functions:
\begin{equation}
\mathcal{T}_{\mu }^{VS}\left( q\right) =\mathcal{T}_{\mu \nu }^{AV}\left(
q\right) =0,
\end{equation}
\begin{eqnarray}
\mathcal{T}^{SS}\left( q\right) &=& 4 \Biggl\{ \left[
I_{quad}\left( M^{2}\right) \right] +\frac{1}{2}\left(
4M^{2}-q^{2}\right) \left[
I_{log}\left( M^{2}\right) \right]   \notag \\
&&- \frac{1}{2}\left( 4M^{2}-q^{2}\right) \left( \frac{i}{16\pi
^{2}}\right) \left[ Z_{0}\left( q^{2},M^{2}\right) \right]
\Biggr\} , \label{T-SS}
\end{eqnarray}
\begin{eqnarray}
\mathcal{T}^{PP}\left( q\right) &=& 4 \Biggl\{ -\left[
I_{quad}\left( M^{2}\right) \right] + \frac{1}{2}q^{2}\left[
I_{log}(M^{2})\right]
\notag \\
&& - \frac{1}{2}q^{2}\left( \frac{i}{16\pi ^{2}}\right) \left[
Z_{0}\left( q^{2},M^{2}\right) \right] \Biggr\} , \label{T-PP}
\end{eqnarray}
\begin{equation}
\mathcal{T}_{\mu }^{PA}\left( q\right) =4Mq_{\mu }\left\{ \left[
I_{log}(M^{2})\right] -\left( \frac{i}{16\pi ^{2}}\right) \left[
Z_{0}\left( q^{2},M^{2}\right) \right] \right\} ,  \label{T-PA}
\end{equation}
\begin{eqnarray}
\mathcal{T}_{\mu \nu }^{VV}\left( q\right) &=&\frac{4}{3}\left( q^{2}g_{\mu
\nu }-q_{\mu }q_{\nu }\right) \notag \\
&&\times \left\{ \left[ I_{log}(M^{2})\right] -\left( \frac{i}{16\pi ^{2}}%
\right) \left[ \frac{1}{3}+\frac{(2M^{2}+q^{2})}{q^{2}}\left[
Z_{0}\left( q^{2},M^{2}\right) \right] \right] \right\} ,
\label{T-VV}
\end{eqnarray}
\begin{eqnarray}
\mathcal{T}_{\mu \nu }^{AA}\left( q\right) &=&\frac{4}{3}\left( q^{2}g_{\mu
\nu } - q_{\mu }q_{\nu }\right)  \notag \\
&&\times \left\{ \left[ I_{log}(M^{2})\right] -\left( \frac{i}{16\pi ^{2}}%
\right) \left[ \frac{1}{3}+\frac{(2M^{2}+q^{2})}{q^{2}}\left[
Z_{0}\left(
q^{2},M^{2}\right) \right] \right] \right\}  \notag \\
&&-8M^{2}g_{\mu \nu }\left\{ \left[ I_{log}(M^{2})\right] -\left( \frac{i}{%
16\pi ^{2}}\right) \left[ Z_{0}\left( q^{2},M^{2}\right) \right]
\right\} . \label{T-AA}
\end{eqnarray}
The tensor amplitudes can be written as
\begin{eqnarray}
\mathcal{T}_{\mu \nu }^{ST}\left( q\right) &=& \mathcal{T}_{\mu
\nu }^{PT} \left( q\right) =\mathcal{T}_{\alpha \mu \nu
}^{AT}\left( q\right) = 0, \\
\mathcal{T}_{\alpha \mu \nu }^{VT}\left( q\right) &=&4M\left(
g_{\alpha \nu }g_{\mu \lambda }-g_{\alpha \mu }g_{\nu \lambda
}\right) q^{\lambda }\left\{ [I_{log}(M^{2})]-\left(
\frac{i}{16\pi ^{2}}\right) \left[ Z_{0}\left( q^{2},M^{2}\right)
\right] \right\} ,  \label{T-VT} \\
\mathcal{T}_{\alpha \beta \mu \nu }^{TT}\left( q\right) &=&\left( g_{\alpha
\mu }g_{\beta \xi }g_{\nu \lambda }-g_{\alpha \nu }g_{\beta \xi }g_{\mu
\lambda }+g_{\beta \nu }g_{\alpha \xi }g_{\mu \lambda }-g_{\beta \mu
}g_{\alpha \xi }g_{\nu \lambda }\right)  \notag \\
&&\times \frac{4}{3}\left( q^{2}g^{\xi \lambda }-q^{\xi
}q^{\lambda }\right) \left\{ I_{log}(M^{2})-\left( \frac{i}{16\pi
^{2}}\right) \left[ \frac{1}{3}+\frac{2M^{2}+q^{2}}{q^{2}}
\left[ Z_{0}\left( M^{2},q^{2}\right) \right] \right] \right\}  \notag \\
&&+ 4\left( g_{\alpha \nu }g_{\beta \mu }-g_{\alpha \mu }g_{\beta
\nu }\right) \Biggl\{ -\left[ I_{quad}(M^{2})\right] +
\frac{1}{2}\left(4M^{2}+q^{2}\right) [I_{log}(M^{2})]  \notag \\
&&- \frac{1}{2}\left( 4M^{2}+q^{2}\right) \left( \frac{i}{16\pi
^{2}}\right) \left[ Z_{0}\left( q^{2},M^{2}\right) \right]
\Biggr\} . \label{T-TT}
\end{eqnarray}

\section{Phenomenology and numerical results}

\label{sec:numerical}

The amplitudes obtained in the preceding section are free of ambiguities and
are symmetry preserving. In this paper we have focused on one and two-point
functions, where resides the highest degree of divergence,
however similar results can be obtained for three and four-point functions used in the 
model to describe meson decays and meson-meson interactions. All these 
mathematical structures also appear in fundamental theories. Here, as a
consequence of the adopted strategy to handle the divergent structures,
it must be noted that such structures are treated in a very closely related 
way as in renormalization procedures of renormalizable theories.
In the obtained expressions for the calculated amplitudes only two
divergent objects have survived after the adoption of CR's, namely 
$I_{log}(M^2)$ and $I_{quad}(M^2)$. The next step, if the amplitudes were to be 
considered in the context of fundamental theories, is the elimination
of such objects through the reparametrization of the theory in the renormalization
of physical parameters.
Since $I_{log}(M^2)$ and $I_{quad}(M^2)$ are completely absorbed in this process,
the regularization eventually used plays no relevant role, due to the fact
that the renormalized expressions are independent of particular aspects
of the chosen regularization.
The theory is predictive, given the fact that the results for the amplitudes
associated to physical processes do not depend on the choices involved in the 
intermediary steps.

In the case of NJL model, considering the nonrenormalizable character, 
the remaining undefined objects need to be specified in some way by 
using physical parameters chosen as inputs of the model.
The traditional way to make predictions  in the context of NJL involves 
regularization such as the introduction of a 3D- or 4D-cutoff $\Lambda$
in the involved Feynman integrals. The expressions for the amplitudes
are, in this way, dependent on the parameters of the chosen regularization 
distribution as well as, if the CR's are not satisfied, on the chosen 
routing for internal lines momenta. In this scenario the model contains 
at least two parameters (in the chiral limit $m_0=0$); the coupling 
strength $G_S$ and a regularization parameter (the cutoff $\Lambda$).
The constituent quark mass $M$ is not an input parameter as it is 
given by the solution of the gap equation. The parameters $G_S$ and 
$\Lambda$ need to be fixed through two experimental information.
The quark condensate $\langle\overline{\psi}\psi\rangle$ and the pion 
decay constant $f_{\pi}$ are usually used for this purpose.
The first is related to a quadratically divergent amplitude while the second
is related to a logarithmically divergent one. The values for $G_S$
and $\Lambda$ which give the best adjustments to the experimental values of
$\langle\overline{\psi}\psi\rangle$ and $f_{\pi}$ will depend on the specific 
form of the regularization distribution. As a consequence all the physical 
amplitudes describing processes pertinent to the model are affected by the 
choice of the regularization. Due to this reason the chosen regularization 
must be part of the model. The predictive power of the original quantum 
field model is affected since the predictions are dependent on a choice which 
characterizes an ambiguity. This is not the desirable situation. 
We wish any model prediction becomes unique in a similar way as it happens 
in renormalizable theories. The regularization must become just a convenient 
choice in the intermediary steps. In order to show that it is possible to achieve
the desirable situation, referred above, we first note that the adoption of the 
CRA's, listed in the preceding Section, implies that we can only adopt regularizations 
that fulfill the CR's. This does not represent a choice as, following our 
analysis, regularizations which break the CR's will lead to ambiguities as well as
to symmetry violations and this is unacceptable. If we want a predictive model, 
only regularizations preserving the CR's make sense .
After these important remarks let us now show how the arbitrariness associated 
to the choice of regularization can be completely removed. 

Having this in mind we start by showing how to relate the remaining objects
$I_{log}(M^2)$ and $I_{quad}(M^2)$ to physical observables chosen as input 
of the model. First we point out that the quark condensate
$\langle\overline{\psi}\psi\rangle$ is related to the $\mathcal{T}^{S}$ one-point
function as 
\begin{equation}
\langle\overline{\psi}\psi\rangle=-N_c\mathcal{T}^{S}.
\end{equation}
Substituting the result (\ref{T-S}) we get
\begin{equation}
\langle\overline{\psi}\psi\rangle=-4N_{c}M\left[  iI_{quad}\left(
M^{2}\right)  \right]  .\label{cond-Iquad}
\end{equation}
For the second we note that, in the context of NJL model, mesons are
relativistic quark-antiquark bound states. In the random-phase approximation
(RPA), the meson propagators can be written as (see for example
Ref.~\cite{Klevansky})
\begin{equation}
D_{\mathcal{M}}\left(  q^{2}\right)  =\frac{2G_{S}}{1-2G_{S}\Pi_{\mathcal{M}%
}\left(  q^{2}\right)  }\,,\label{m-propagator}%
\end{equation}
where $\Pi_{\mathcal{M}}$ is the polarization function defined by
\begin{equation}
\Pi_{\mathcal{M}}\left(  q^{2}\right)  =i\int\frac{d^{4}k}{\left(
2\pi\right)  ^{4}}\mathrm{Tr}\left\{  \Gamma_{\mathcal{M}}S\left(
k+k_{1}\right)  \Gamma_{\mathcal{M}}S\left(  k+k_{2}\right)  \right\}  ,
\end{equation}
with $S$ being the quark propagator defined previously. 
$\Gamma_{\mathcal{M}}$ stands for the flavor and Dirac matrices giving the
quantum numbers of the meson $\mathcal{M}$. For example, for the neutral pion,
$\Gamma_{\mathcal{M}}=\tau_{3}\gamma_{5}$, for the scalar-isoscalar meson,
$\Gamma_{\mathcal{M}}=1$. In writing the equations above, we assumed the most
general labels for the momenta $k_{1}$ and $k_{2}$ running in the internal
lines of the loop integral. The physical momentum $q$ is defined as the
difference $k_{1}-k_{2}$ as imposed by energy-momentum conservation at each vertex.

The pole of the propagator in Eq.~(\ref{m-propagator}), calculated at
$q^{2}=m_{\mathcal{M}}^{2}$, gives the mass of the respective meson. The
condition for the pion mass is given by
\begin{equation}
1-2G_{S}\Pi_{\pi}(m_{\pi}^{2})=0\,,
\end{equation}
where
\begin{equation}
\Pi_{\pi}\left(  q^{2}\right)  =-iN_{c}N_{f}\left[  \mathcal{T}^{PP}\left(
q\right)  \right]  .
\end{equation}
Substituting now the explicit form of $\mathcal{T}^{PP}\left(  q\right)$,
derived previously and given in Eq.~(\ref{T-PP}), we obtain
\begin{align}
\Pi_{\pi}\left(  q^{2}\right)   &  =4iN_{c}N_{f}\left[  I_{quad}\left(
M^{2}\right)  \right] \nonumber\\
&  -2iN_{c}N_{f}q^{2}\left\{  \left[  I_{log}\left(  M^{2}\right)  \right]
-\left(  \frac{i}{16\pi^{2}}\right)  \left[  Z_{0}\left(  q^{2},M^{2}\right)
\right]  \right\}  . \label{polarization-PP}
\end{align}
Using the Eq.~(\ref{gap1}) in order to eliminate $I_{quad}\left(  M^{2}\right)$
and evaluating Eq.~(\ref{polarization-PP}) at $q^{2}=m_{\pi}^{2}$ we get the
following expression for the pion mass
\begin{equation}
m_{\pi}^{2}=-\frac{m_{0}}{4N_{c}N_{f}MG_{S}}\frac{1}{\left\{  i\left[
I_{log}\left(  M^{2}\right)  \right]  +\left(  \frac{1}{16\pi^{2}}\right)
\left[  Z_{0}\left(  m_{\pi}^{2},M^{2}\right)  \right]  \right\}  }.
\label{mass-pion1}
\end{equation}
As seen, in the chiral limit ($m_{0}=0$) the pion becomes massless $\left(
m_{\pi}=0\right)  $, in agreement with Goldstone's theorem.

The pion phenomenology is also characterized by the decay constant $f_{\pi}$.
Experimentally it is related to the weak decay $\pi^{\pm}\rightarrow\mu^{\pm
}+\nu_{\mu}$ and is calculated from the vacuum to one-pion axial-vector
current matrix element
\begin{equation}
\langle0|\overline{\psi}(x)\gamma_{\mu}\gamma_{5}\tau^{i}/2\psi(x)|\pi
^{j}\left(  q\right)  \rangle=i\ f_{\pi}q_{\mu}\delta_{ij}e^{-iqx},
\end{equation}
where $|\pi^{j}(q)\rangle$ is a pion state with four-momentum $q$. At one-loop
order, one can express this matrix element in terms of the $\mathcal{T}_{\mu
}^{AP}$ two-point function as
\begin{equation}
if_{\pi}q_{\mu}\delta_{ij}=-N_{c}g_{\pi qq}\delta_{ij}\left[  \mathcal{T}
_{\mu}^{AP}\left(  q\right)  \right]  ,
\end{equation}
where $g_{\pi qq}$ is the pion-to-quark-quark coupling strength, related to
the residue of Eq.~(\ref{m-propagator}) as
\begin{equation}
g_{\pi qq}^{2}=\left.  \left(  \frac{\partial\Pi_{PP}\left(  q^{2}\right)
}{\partial q^{2}}\right)  ^{-1}\right\vert _{q^{2}=m_{\pi}^{2}}.
\end{equation}
Using Eqs.~(\ref{T-PA}) and (\ref{polarization-PP}), we can write
\begin{align}
f_{\pi}  &  =-4N_{c}g_{\pi qq}M\left\{  i\left[  I_{log}\left(  M^{2}\right)
\right]  +\left(  \frac{1}{16\pi^{2}}\right)  \left[  Z_{0}\left(  m_{\pi}
^{2},M^{2}\right)  \right]  \right\}  ,\label{f-pion1}\\
g_{\pi qq}^{-2}  &  =-2N_{c}N_{f}\left\{  i\left[  I_{log}(M^{2})\right]
+\left(  \frac{1}{16\pi^{2}}\right)  \left[  Z_{0}\left(  m_{\pi}^{2}
,M^{2}\right)  \right]  \right\} \nonumber\\
&  -2N_{c}N_{f}m_{\pi}^{2}\left(  \frac{1}{16\pi^{2}}\right)  \left[
Y_{1}\left(  m_{\pi}^{2},M^{2}\right)  \right]  \label{g-pion1}
\end{align}
where $Y_{1}\left(q^{2},M^{2}\right)  $ is
the $k=1$ element of the set
\begin{equation}
Y_{k}\left(  q^{2};m^{2}\right)  =\int_{0}^{1}dz\frac{z^{k}\left(  1-z\right)
}{ q^{2}z\left(  1-z\right)  -m^{2}} . \label{definition-Y1}
\end{equation}
In a completely similar way, for the scalar meson ($\sigma$) we have
\begin{align}
m_{\sigma}^{2}  &  =4M^{2}-\frac{m_{0}}{M}\frac{1}{4G_{S}N_{c}N_{f}\left\{
iI_{log}(M^{2})+\left(  \frac{1}{16\pi^{2}}\right)  \left[  Z_{0}\left(
m_{\sigma}^{2},M^{2}\right)  \right]  \right\}  },\\
g_{\sigma qq}^{-2}  &  =2iN_{c}N_{f}\left(  m_{\sigma}^{2}-4M^{2}\right)
\left(  \frac{i}{16\pi^{2}}\right)  \left[  Y_{1}\left(  m_{\sigma}^{2}
,M^{2}\right)  \right] \nonumber\\
&  -2iN_{c}N_{f}\left\{  I_{log}(M^{2})-\left(  \frac{i}{16\pi^{-2}}\right)
\left[  Z_{0}\left(  m_{\sigma}^{2},M^{2}\right)  \right]  \right\}  .
\end{align}

It is easy to see from Eqs.~(\ref{f-pion1}) and (\ref{g-pion1}) that the observable 
$f_{\pi}$ can be related to the undefined quantity $I_{log}$. 
In order to make the aspects we want to emphasize clear, we initially 
consider this relation in the chirally symmetric case.
Then we get
\begin{equation}
iI_{log}\left(  M^{2}\right)  =-\frac{f_{\pi}^{2}}{2N_{c}N_{f}M^{2}
},\label{I-log2}
\end{equation}
since $Z_{0}(m_{\pi}^{2}=0)=0$. 
Through the Eqs.~(\ref{cond-Iquad}) and (\ref{I-log2}) we have 
stated relations between two observables (inputs), the quark condensate and the
pion decay constant, with two undefined objects coming from loop calculations.
These objects are functions of the constituent quark mass which must be 
determined in some stage. Therefore the Eqs.~(\ref{cond-Iquad}) and (\ref{I-log2})
in fact represent the relation of two quantities to two functions.
If we want to know such functions we have to integrate
$I_{log}(M^2)$ and $I_{quad}(M^2)$ which means to adopt an explicit form of 
regularization distribution. This process introduces at least one regularization
parameter $\Lambda$ as it is well known. Different regularizations will generally lead 
to different values of $\Lambda$ as it is usual in the context of NJL model with 
regularizations. In order to avoid this situation we will proceed in
a different way.

First we note that those two functions are not independent. It is possible
to show that they are related by 
\begin{equation}
\frac{\partial}{\partial M^{2}}\left[  i\,I_{quad}\left(  M^{2}\right)
\right]  =i\,I_{\log}\left(  M^{2}\right).\label{rel1}
\end{equation}
On the other hand, we can also state that $I_{\log}(M^2)$ possesses the
following property
\begin{equation}
\frac{\partial}{\partial M^{2}}\left[  i\,I_{\log}\left(  M^{2}\right)
\right]  =\frac{1}{16\pi^{2}M^{2}}.\label{rel2}
\end{equation}
In order to satisfy these two conditions it is necessary to get the following 
general forms
\begin{align}
i\,I_{\log}\left(  M^{2}\right)   &  =\frac{1}{16\pi^{2}}\ln M^{2}+C_{1},\\
i\,I_{quad}\left(  M^{2}\right)   &  =\frac{1}{16\pi^{2}}M^{2}\left[  \ln
M^{2}-1+C_{1}\right]  + C_{2},
\end{align}
where $C_{1}$ and $C_{2}$ are indeterminate constants - $C_{1}$ is dimensionless
and $C_{2}$ has dimension of (mass)$^{2}$. In the context of regularizations,
$C_{1}$ and $C_{2}$ are related to the regularization parameter $\Lambda$.
Eliminating the constant $C_{1}$ we see that
\begin{equation}
i\,I_{quad}\left(  M^{2}\right)  =\frac{-1}{16\pi^{2}}M^{2}+M^{2}\left[
i\,I_{\log}\left(  M^{2}\right)  \right]  +C_{2}.\label{rel3}%
\end{equation}
Replacing $I_{quad}$ and $I_{log}$ in term of $\langle\bar{\psi}\psi\rangle$
and $f_{\pi}$, Eqs. (\ref{I-log2}) and (\ref{cond-Iquad}), we get
\begin{equation}
\frac{M^{3}}{16\pi^{2}}+\left(  \frac{f_{\pi}^{2}}{2N_{c}N_{f}}-C_{2}\right)
M-\frac{\left\langle \overline{\psi}\psi\right\rangle }{4N_{c}}=0.\label{rel4}
\end{equation}
There are two important aspects involved in the above equation. 
First, it ensures that it is crucial to obey the properties 
(\ref{rel1}) and (\ref{rel2})
when the functions $I_{log}(M^2)$ and $I_{quad}(M^2)$ are made
explicit. These two properties work as additional constraints 
to be required of a regularization distribution if one wants 
to get consistency in perturbative calculations. 
The violation of these properties will result in breaking
the scale properties of physical amplitudes, and
possesses the same status of symmetry violations \cite{Orimar-scale}.
The second aspect refers
to the dependence of the physical parameters on the choice of the
specific regularization. The above expression states that even if 
a regularization obeys the CR's and simultaneously the conditions
(\ref{rel1}) and (\ref{rel2}) it remains a freedom to distinguish 
it from other regularizations belonging
to the class of consistent regularizations, which is the value of $C_2$.
Two consistent regularizations can differ only by the value of $C_2$.
As a consequence, the physical implications of the model seem to
be definitely regularization dependent. The introduction of experimental values for the
inputs $\langle\overline{\psi}\psi\rangle$ and $f_{\pi}$, it makes
necessary to specify $C_2$ to get $M$ and then through
the gap equation, to get the value for $G_S$.
There is nothing more to be imposed, based on consistency reasons, 
to remove this arbitrariness. Apparently all our efforts cannot 
avoid the dependence of the results on the choice 
of the regularization, even if we have drastically restricted the 
regularizations which can be used in the calculations.

At this point it seems there is nothing else to do except to choose convenient values 
for $C_2$. The convenience of such choices is related to the fact that we need to choose
an adequate value for
$C_2$ in order to get a good value for $M$ and after this to find a good value for $G_S$. 
It is precisely in this process that emerges
the most surprising aspect of our formulation.
If we recognize that only positive values of $M$ make sense in the
equation, the nonlinear character of the equation in $M$ produces
a critical condition to possible values of $C_2$.  
It turns out that one finds
solutions with $M>0$ only for $C_{2}\geq C_{crit}$, such that
\begin{equation}
\begin{array}
[c]{c}
C_{2}<C_{crit}\rightarrow\text{no solutions}\\
C_{2}=C_{crit}\rightarrow\text{one solution}\\
C_{2}>C_{crit}\rightarrow\text{two solutions}.
\end{array}
\end{equation}
Therefore, it seems obvious that there is only one value for $C_{2}$
which is reasonable, the $C_{crit}$, due to the fact that only this value 
allows us a consistent physical interpretation of the model predictions.
Assuming this attitude, the NJL model becomes predictive since the remaining
arbitrariness is fixed through the existence of a critical condition.
There are considerable differences between our formulation and the traditional ones.
The first and immediate refers to the determination of the constituent
quark mass $M$.
Such value is fixed by the critical point in the diagram $M\times C_2$.
The value of the mass, therefore, depends only on the experimental values for  
$\left\langle \overline{\psi}\psi\right\rangle$ and $f_{\pi}$ which are the
chosen inputs of the model. The gap equation will be used in the determination
of the coupling $G_S$ compatible with the mass fixed by the critical condition.
At this point it is crucial to ask: are the values for $M$ and $G_S$,
emerging from this critical condition, reasonable?

\begin{figure}[h]
\includegraphics[width=230pt,height=150pt]{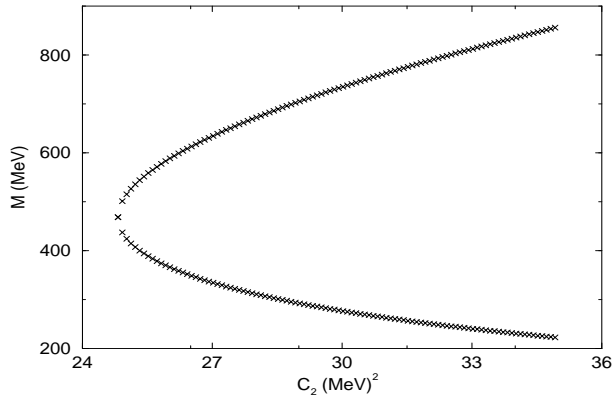}\caption{Solutions of
Eq.(\ref{rel4}) as function of the constant $C_{2}$.}
\label{critical}
\end{figure}

In Fig.~\ref{critical} we plot all possible physical solutions ($M>0$) of
Eq.~(\ref{rel4}) as function of the arbitrary constant~$C_{2}$.
Using as input $\left\langle \overline{\psi}\psi\right\rangle = \left(
-250.0\;\mathrm{MeV}\right)  ^{3}$ and $f_{\pi} = 93.0\;\mathrm{MeV}$ it results
that for values of $C_2 < C_{crit}$, Eq.~(\ref{rel4}) there is no physical
solution, at $C_{2}=C_{crit}$ there is only one solution, and for $C_{2} >
C_{crit}$ there are two possible solutions. In particular, at the critical
point we obtain $C_{crit}\simeq 24.82\;\mathrm{MeV}^{2}$, $M\simeq
468.4\;\mathrm{MeV}$ and $G_{S}\simeq 7.5\;\mathrm{GeV}^{-2}$.
Therefore the values for $M$ and $G_S$ are in good agreement with the ones
found in literature of this issue. 

Another difference between this formulation and the traditional ones
refers to the meson phenomenology. When we adopt the present formulation,
in the presence of chiral symmetry breaking parameter $m_0$, previously
fixed, since it is an input parameter, the meson masses
and meson-quark-quark couplings, as well as other physical aspects,
emerge as genuine predictions, including those of pion. In order to
see this aspect clearly let us consider the case with $m_{0} \neq0$. 
As it can be seen
in Eq.~(\ref{mass-pion1}), the pion mass is nonzero, reflecting the fact that
the original Lagrangian is not chirally symmetric with $m_{0} \neq0$. The
introduction of a current quark mass modifies the results for other quantities
as well. For example, the expression for $f_{\pi}$ now contains a finite part,
and because of this, the expression of $I_{log}\left(  M^{2}\right)  $ in
terms of $f^{2}_{\pi}$ changes to
\begin{align}
i\left[  I_{log}\left(  M^{2}\right)  \right]   &  =-\frac{\left[
Z_{0}\left(  m_{\pi}^{2},M^{2}\right)  \right]  }{16\pi^{2}}\nonumber\\
&  -\frac{f_{\pi}^{2}}{4N_{c}N_{f}M^{2}}\left(  1+\sqrt{1-\frac{ N_{c}%
N_{f}m_{\pi}^{2}M^{2}\left[  Y_{1}\left(  m_{\pi}^{2},M^{2}\right)  \right]
}{2\pi^{2}f_{\pi}^{2}}}\right).
\end{align}
As a result, we can write for the pion mass the following expression
\begin{equation}
m_{\pi}^{2}=\frac{m_{0}M}{f_{\pi}^{2}G_{S}}\left(  1+\sqrt{1-\frac{ N_{c}
N_{f}m_{\pi}^{2}M^{2}\left[  Y_{1}\left(  m_{\pi}^{2},M^{2}\right)  \right]
}{2\pi^{2}f_{\pi}^{2}}}\right)  ^{-1}. \label{mass-pion2}
\end{equation}
Also, for the effective pion-coupling constant we obtain
\begin{align}
g_{\pi qq}^{-2}  &  =-\frac{N_{c}N_{f}m_{\pi}^{2}}{8\pi^{2}}\left[
Y_{1}\left(  m_{\pi}^{2},M^{2}\right)  \right] \nonumber\\
&  +\frac{f_{\pi}^{2}}{2M^{2}}\left(  1+\sqrt{1-\frac{N_{c}N_{f}m_{\pi}
^{2}M^{2}\left[  Y_{1}\left(  m_{\pi}^{2},M^{2}\right)  \right]  }{2\pi
^{2}f_{\pi}^{2}}}\right)  . \label{g-pion3}
\end{align}
As before, eliminating $I_{quad}(M^{2})$ and $I_{log}(M^{2})$
in favor of $\langle\bar\psi\psi\rangle$ and $f_{\pi}$, we obtain a nonlinear
equation for $M$ that now involves the pion mass $m_{\pi}$,
\begin{align}
\frac{M^{3}}{16\pi^{2}}\left[  1+Z_{0}\left(  m_{\pi}^{2},M^{2}\right)
\right]   &  =\frac{\left\langle \overline{\psi}\psi\right\rangle }{4N_{c}}
+C_{2}M\nonumber\\
&  -\frac{f_{\pi}^{2}M}{4N_{c}N_{f}}\left(  1+\sqrt{1-\frac{N_{c}N_{f}m_{\pi
}^{2}M^{2}\left[  Y_{1}\left(  m_{\pi}^{2},M^{2}\right)  \right]  }{2\pi
^{2}f_{\pi}^{2}}}\right)  . \label{rel5}
\end{align}
This equation is much more complicated to solve the one in the case of exact
symmetry. This equation and the expression for $m_{\pi}$, however, can be
simplified using the following approximations for $Z_{0}$ and $Y_{1}$
\begin{equation}
Z_{0}\left(  q^{2};m^{2}\right)  =-\frac{q^{2}}{6m^{2}}-\frac{q^{4}}{60m^{4}}
+\cdots, \label{approx-Z0}
\end{equation}
and
\begin{equation}
Y_{1}\left(  q^{2};m^{2}\right)  =-\frac{1}{6m^{2}}-\frac{q^{2}}{30m^{4}}
+\cdots. \label{approx-Y1}
\end{equation}
Using these approximations, we obtain
\begin{align}
&  m_{\pi}^{2}=\frac{m_{0}M}{f_{\pi}^{2}G_{S}}\left(  1+\sqrt{1+\frac{
N_{c}N_{f}m_{\pi}^{2}}{12\pi^{2}f_{\pi}^{2}}}\right)  ^{-1},\label{mass-pion3}
\\
&  \frac{M^{3}}{16\pi^{2}}+\frac{f_{\pi}^{2}M}{4N_{c}N_{f}}\left(1 + \sqrt{1 +
\frac{N_{c}N_{f}m_{\pi}^{2}}{12\pi^{2}f_{\pi}^{2}}}\right)  -\frac{m_{\pi}
^{2}M}{96\pi^{2}}-C_{2}M-\frac{\left\langle \overline{\psi}\psi\right\rangle
}{4N_{c}}=0. \label{rel6}
\end{align}
Inserting the ``experimental'' value for $m_0$ (as well as for 
$\left\langle \overline{\psi}\psi\right\rangle$ and $f_{\pi}$) we search for the
value of $C_2$ which corresponds to only one positive value for $M$ and thus
determine the values for $G_{S}$, $m_{\pi}$, and so on. For this propose we
take $m_{0} = 5.2\;\mathrm{MeV}$, obtaining $m_{\pi}\simeq 135.3\;\mathrm{MeV}$,
$M\simeq468.4\;\mathrm{MeV}$, $G_{S}\simeq 7.5\;\mathrm{GeV}^{-2}$ 
and $C_{crit}\simeq 24.82\;\mathrm{MeV}^{2}$.
Finally, Eq.(\ref{g-pion3}) furnish $g_{\pi qq}\simeq 4.97$. 
These predictions are in good agreement with experimental data and those used 
in the literature.

In the scalar channel, we have the following expressions for the $\sigma$
meson mass and the $\sigma qq$ coupling constant
\begin{align}
m_{\sigma}^{2} &  =4M^{2}-\frac{m_{0}}{MG_{S}}\left[  \frac{N_{c}N_{f}}
{4\pi^{2}}\left[  Z_{0}\left(  m_{\sigma}^{2},M^{2}\right)  +\frac{m_{\pi}
^{2}}{6M^{2}}\right]  \right.  \nonumber\\
&  \left.  -\frac{f_{\pi}^{2}}{M^{2}}\left(  1+\sqrt{1+\frac{N_{c}N_{f}m_{\pi
}^{2}}{12\pi^{2}f_{\pi}^{2}}}\right)  \right]  ^{-1},\\
g_{\sigma qq}^{-2} &  =-\frac{N_{c}N_{f}}{8\pi^{2}}\left[  Z_{0}\left(
m_{\sigma}^{2},M^{2}\right)  +\frac{m_{\pi}^{2}}{6M^{2}}\right]  \nonumber\\
&  -\frac{N_{c}N_{f}}{8\pi^{2}}\left(  m_{\sigma}^{2}-4M^{2}\right)  \left[
Y_{1}\left(  m_{\sigma}^{2},M^{2}\right)  \right]  \nonumber\\
&  +\frac{f_{\pi}^{2}}{2M^{2}}\left(  1+\sqrt{1+\frac{N_{c}N_{f}m_{\pi}^{2}
}{12\pi^{2}f_{\pi}^{2}}}\right)  .
\end{align}
Numerically we have $m_{\sigma}=938\;\mathrm{MeV}$ and $g_{\sigma qq}=2.29$.

The fact of the arbitrary character of $C_{2}$ being removed, owing to the 
existence of a critical condition, it is the most important result point in the
analysis made in this Section. This means that the phenomenology becomes completely
independent of the specific regularization scheme employed if such scheme is
consistent with the scale invariance. 

The model, within the scope of this prescription, becomes predictive since the 
role played by a regularization has completely disappeared. In this sense, 
in spite of being a nonrenormalizable model, the predictions are made in the 
same spirit as in renormalized models, at the considered level of approximation. 
This is, undoubtedly, a very important improvement in the quality of this type of 
perturbative calculations. However, we must be aware of the fact that this does not 
represent the solution to all the problems involved. Since the amplitudes have acquired 
structures which are very similar to those belonging to the renormalized theories, what 
remains to be considered are the phenomenological implications for the model predictions 
of the so called Landau vacuum instabilities or ghost poles in boson propagators. 
This aspect of the perturbative calculations appeared in hadron physics in connection 
with the Skyrme model \cite{Ripka-kahana, Sony, Banerjee, Meissner-Arriola}. 
More specifically, Landau vacuum instabilities were found in chiral-quark models when 
soliton solutions were searched for in the renormalized sigma model. 
(For an update of this problem please see the recent paper of Arriola-Broniowski-Golli 
\cite{Arriola-Broniowski}). However, Landau instabilities seem to be present in almost 
all renormalizable theories where fermions are coupled with boson fields \cite{Banerjee}. 
Their occurrence can be indicated by the presence of tachyon poles in boson propagators 
corrected by one-loop fermionic contributions \cite{Perry}. In asymptotically free theories 
where the bosons are self-interacting fields, contributions coming from bosonic one-loop 
diagrams may eliminate the problem \cite{Perry}, otherwise the general rule seems to be the 
existence of vacuum instabilities \cite{Banerjee}. In the QED, where the vector gauge field 
is not a self-interacting field, tachyonic poles occur in the photon propagator such that 
the vacuum is unstable at the one-loop level. In the linear sigma model the meson propagators 
are equally contaminated by tachyonic poles. In the QED, the presence of such type of undesirable 
poles does not play a physically relevant role because the scale of the fluctuations at which 
the instabilities occur is $m\,e^\frac{1}{\alpha}$ ($m$ is the electron mass 
and $\alpha$ is the fine structure constant) \cite{Banerjee}. Therefore, this happens in a 
region which is certainly beyond the expected validity of the theory. On the other hand, 
in hadronic phenomenological models, where a fermionic field is coupled with a mesonic one, 
like in the Yukawa model or in the chiral $\sigma$ model, such scale of flutuations changes 
drastically (around 1GeV) due to the nucleon or quark mass and due to the value of the constant 
coupling involved. This means that the ghost poles may have relevant influence in phenomenological 
implications of the model.

In NJL model we have only fermions in the Lagrangian but the meson states are interpreted 
as a quark-antiquark bound states. The intermediate amplitudes are fermionic loops such that, 
in the RPA approximation, the meson mass is identified as the pole of the Eq. (107). 
This is precisely the structure of the renormalized meson propagator in the linear $\sigma$ model. 
This means that if we look carefully to the condition stating the pion mass, for example, we will 
find that in addition to the pion pole, there is a tachyonic pole in the Euclidian region with a 
negative residue. This implies that the corresponding dispersion relation will be verified only 
if this pole is included. This situation is not commonly considered within the context of NJL model 
due to the use of cut-offs which change the behavior of the fermionic one-loop contributions such 
that the problem is automatically eradicated \cite{Arriola-Broniowski}.

In the procedure adopted in the present work the finite parts of the Green's functions are not modified, 
putting the physical amplitudes at the same level as those belonging to renormalizable theories. 
Because of this, the questions related to the ghost poles or vacuum instabilities may become relevant. 
The presence of ghost poles in meson propagators or Landau vacuum instabilities in the NJL model 
is expected due to its equivalence to the linear $\sigma$ model. Although the Green's functions are, 
strictly speaking, not the same ones, due to the definition of the renormalization parameters, 
the $S$-matrices of both theories are identical \cite{Eguchi}.

On general grounds it is not completely clarified if ghost poles are real ingredients of 
QFT or if they are a product of a particular kind of perturbative solution (one-loop approximation) \cite{Banerjee}. 
Due to the fact that if they are real ingredients of QFT fundamental axioms are violated since only 
real poles are expected to exist, the relevant question seems to investigate if this type of instabilities 
survives to higher-orders calculations. The acceptance that these undesirable poles are unavoidable and, 
therefore, real aspects of QFT, constitutes a very frustrating fact just because we are accepting that 
the solutions we can obtain do not obey the fundamental axioms of the theoretical apparatus we have constructed. 
The most reasonable expectation is that the instabilities will disappear when contributions of higher-order are 
computed. If this is the case, there is nothing else to do but concluding that the one-loop approximations are 
not adequate to investigate phenomenological implications of a theory or model. If, however, the instabilities 
constitute an unavoidable aspect of certain classes of QFT, we must make efforts to get theories free from those 
problems by construction, like in the case of anomalies. Only additional investigations will clarify these doubts. 
We are, however, convinced that the questions related to the regularizations in QFT are of a different nature 
from those related to the ghost poles in perturbative corrections of boson propagators.

\section{Conclusions and Perspectives}
\label{sec:conclusions}

We considered in detail questions relative to the predictive power of the 
NJL model. Given its non-renormalizability, the model predictions are usually compromising 
with the regularization method employed. The regularization cannot be removed from 
the results and consequently, the physical implications are crucially dependent on 
the adopted regularization technique. It is usual practice consider the regularization 
as part of the model -- see for example~\cite{Klevansky,HK-rev,Bijnens}. However,
depending on the adopted regularization method, physical amplitudes may emerge from 
the calculations ambiguous and symmetry violating. Undefined quantities arising from 
divergences are fixed in the parametrization of the model by adjustments in regularization 
parameters. In veiw of this, one has the potential problem that the results of 
calculations are not real predictions, but particular choices for the involved 
arbitrariness (ambiguities)~\cite{Willey,Gherghetta} that lead to results that
might change when different choices are made.

Our investigation in the present paper focused on avoiding as much as possible explicit 
evaluations of divergent quantities. Our discussion consisted in basically two steps. 
In the first step we obtained physical amplitudes free from ambiguities and symmetry 
preserving. Integrands of divergent Feynman integrals were manipulated such that
all the dependence on internal (arbitrary) momenta is left in terms that lead
to finite integrals, which are then integrated free from the regularizations effects, 
while purely divergent objects (combinations of divergent integrals) can be clearly 
identified. Invoking symmetry constraints and demanding elimination of ambiguities 
lead to what we called consistency relations (CR), which require definite values 
for the divergent objects. All one- and two-point functions of the model, in which 
the highest degrees of divergence reside, emerge from the calculations free from
ambiguities and respecting symmetry constraints. In a second step, a parametrization 
was introduced to eliminate divergences. We considered a parametrization of the 
model where the remaining divergent integrals $I_{quad}\left( M^{2}\right)$ and 
$I_{\log }\left( M^{2}\right)$, need to be eliminated by fixing phenomenological 
quantities. We used general scale properties of $I_{quad}\left(M^{2}\right)$ and 
$I_{\log }\left( M^{2}\right)$ in order to make explicit the freedom one always 
has when choosing a specific regularization through the constant parameter 
$C_{2}$. All this was done without explicit evaluation of the divergent integrals. 

Of course, we could also have chosen a specific regularization for evaluating
explicitly $I_{quad}\left( M^{2}\right) $ and $I_{\log }\left( M^{2}\right)$ and
put our calculation in close connection with the traditional regularization methods.
As a result, both  $I_{quad}\left( M^{2}\right)$ and $I_{\log }\left( M^{2}\right)$
become a function of a regularization parameter $\Lambda $ and would also
depend on the form chosen for the regularization function $G(k^2/\Lambda^2)$.
In this way, for different regularizations one would have different values for
$C_2$. We have shown that without making any choice one can fix this $C_2$ 
by simply choosing its critical value. An important point to be noted, we reiterate,
is that all manipulations done prior to such a choice have been made guarantee 
amplitudes that are free from loop momenta ambiguities and that preserve the 
symmetries of the model. 

The assertion that the formulation of the regularization of the NJL model 
presented here is predictive is to be understood in the sense that no arbitrary 
choices were made in intermediate steps up the stage of calculating 
phenomenological quantities, like the pion and sigma masses and coupling constants. 
Even at this last stage, the existence of a critical value for $C_2$ that leads to good 
values for phenomenological quantities is gratifying. 

There is a variety of applications that can be made using the general formulation 
presented here. Like the recent application to clustered hadronic matter~\cite{matter} 
and color superconductivity in high density quark matter~\cite{super}, we envisage great 
potential for the study of heavy flavor in quark matter. The SU(3) version of NJL
model constitutes a natural candidate to the application of the formulation 
presented here. Work along this line is presently under way \cite{OB_su3}.

\begin{acknowledgments}
Work partially supported by the Brazilian agencies CNPq and FAPESP. 
\end{acknowledgments}

\end{document}